# Comet C/2011 L4 (PanStarrs)

Small nucleus, fast rotator and dust rich comet observed after perihelion

*T. Scarmato[1]*

[1]Toni Scarmato's Observatory, via G. Garibaldi 46, 89817 San Costantino di Briatico, Calabria, Italy

(toniscarmato@hotmail.it; Cellphone:+393479167369; Tel/Fax:+390963392102)

**August 2016 version1.0**

**Abstract**

Orbital elements of **C/2011 L4 (PanStarrs)** Oort cloud comet, computed by MPC (Minor Planet Center, *Minor Planet Electronic Circular 2012–T08*), show that the closest approach to the Sun occurred on **2013 March 10$^{th}$**, at about **0.3 A.U.**, then about **4,51x10^7 km**.

```
C/2011 L4 (PANSTARRS)     Orbital elements by G. V. Williams
          Epoch 2013 Mar. 9.0 TT = JDT 2456360.5
                   T 2013 Mar. 10.16839 TT
          q   0.3015433     (2000.0)       P          Q
          z  -0.0000420   Peri. 333.65160  +0.41006823  +0.10046864
          +/-0.0000009   Node  65.66583   +0.90783024  +0.05059039

          e  1.0000127    Incl. 84.20692  -0.08768299  +0.99365319
```

Discovered by Richard Wainscoat (Institute for Astronomy, University of Hawaii) on four CCD images taken with the 1.8-m "Pan-STARRS 1" telescope at Haleakala taken on 2011 June 6th. My first observation of the comet was on 2013 March 10th whit the comet visible in the twilight. I did the following visual estimation; **Mar. 10.73,-1.0\*,5'(T. Scarmato, Calabria, Italy, 7x50 binoculars; altitude 7 deg, tail 1 deg in pa 140 deg),** reported to the ICQ (International Comet Quarterly)**.** Easy comet in 7x50 binoculars I started to see C/2011 L4 at 18:35 L.T. when the comet was at about 7° above the horizon. I saw a tail long about **1,5° in pa 140°** with a coma well condensed and **large about 5'**. I followed the comet until to the set at **18:55 L.T.**, still clearly visible at so low altitude about 1° or less! I don't saw Mars (1.39 mag) at the same altitude, so I esteemed the comet using Aldebaran (1.1 mag) at 62,44° of altitude also using the **ICQ Table of Atmospheric Extinction**; but remembering C/2006 P1 in the same conditions of observation I could to assume that the comet was at negative apparent total magnitude **m1= -1.0**. The image of the comet was impressive (**see Fig. 2**).

Here, I present my observations and results on the size of the nucleus, period of rotation, dust production and peculiar structures in the inner coma.

## 1) INTRODUCTION

### 1.1 Perihelion passage

The orbit of C/2011 L4 put the comet on a good view for observers in the northern hemisphere from the second week of March 2013.

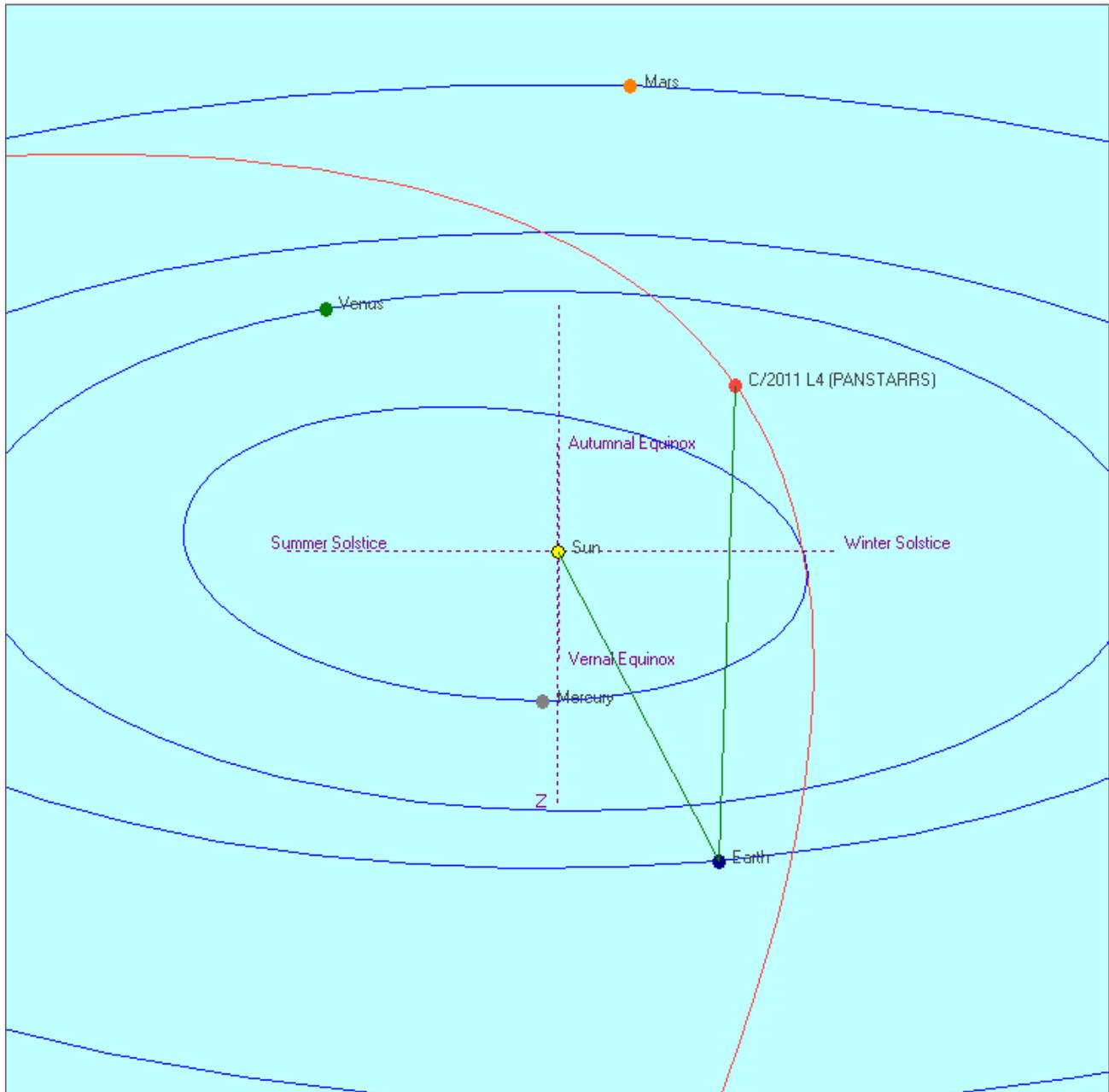

**Fig. 1 – Orbit and position of the comet C/2011 L4 (PanStarrs) on 2013 March 9th plotted to put in evidence the starting of the good observational period for northern observers, between the perihelion time and 2013 summer.**

I tried to observe the comet on **2013 March 9th** without success. On the afternoon of **2013 March 10th**, I pointed my camera Canon 10D on tripod, without AR motor, and with my big surprise I saw the comet in a single shot of 1 second of exposure. **(See following Fig. 2)**

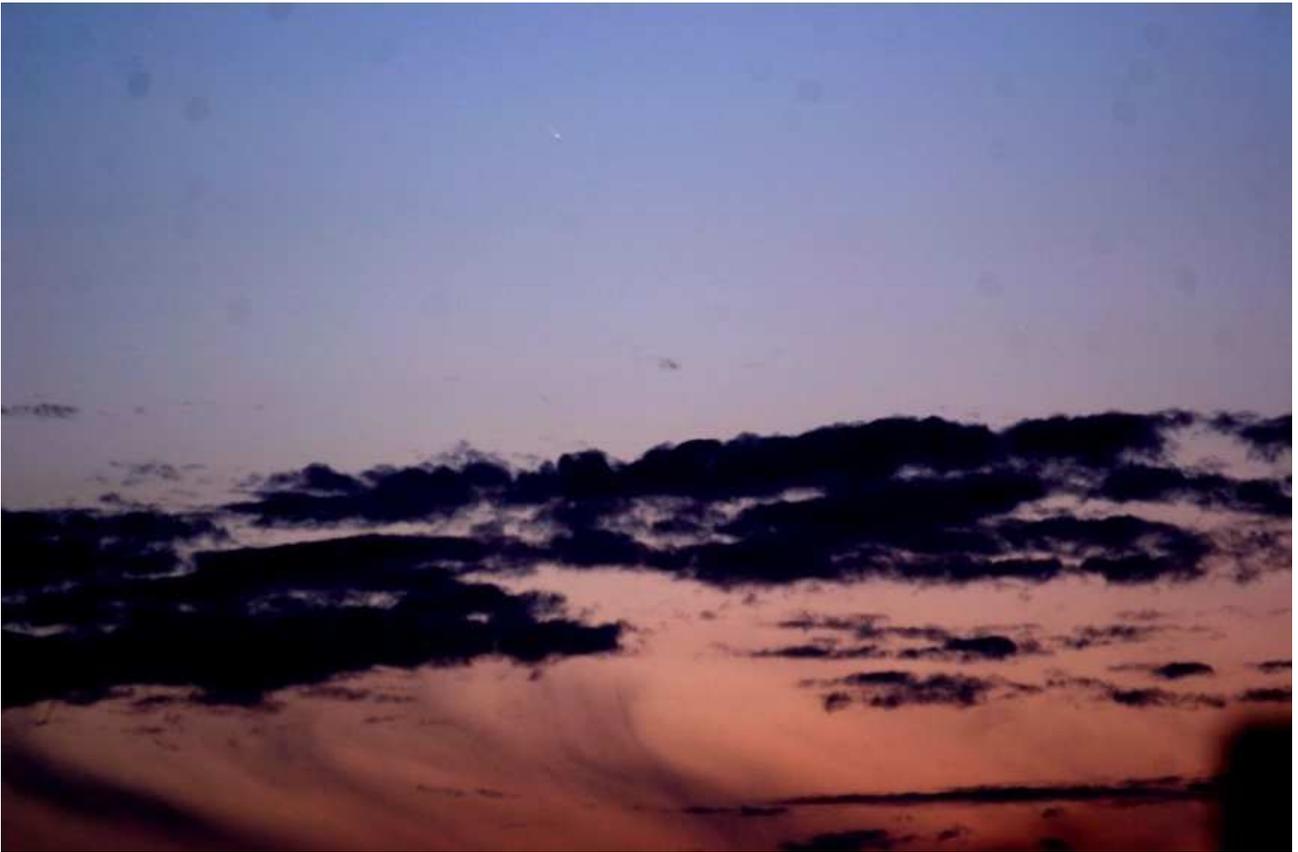

**Fig. 2 – Comet C/2011 L4 (PanStarrs) succesful imaged in the twilight on 2013 March 10 with a Canon 10 D and Tele Apo 300 mm with 1 second of exposure. The comet was observed also visually in 7x50 binoculars.**

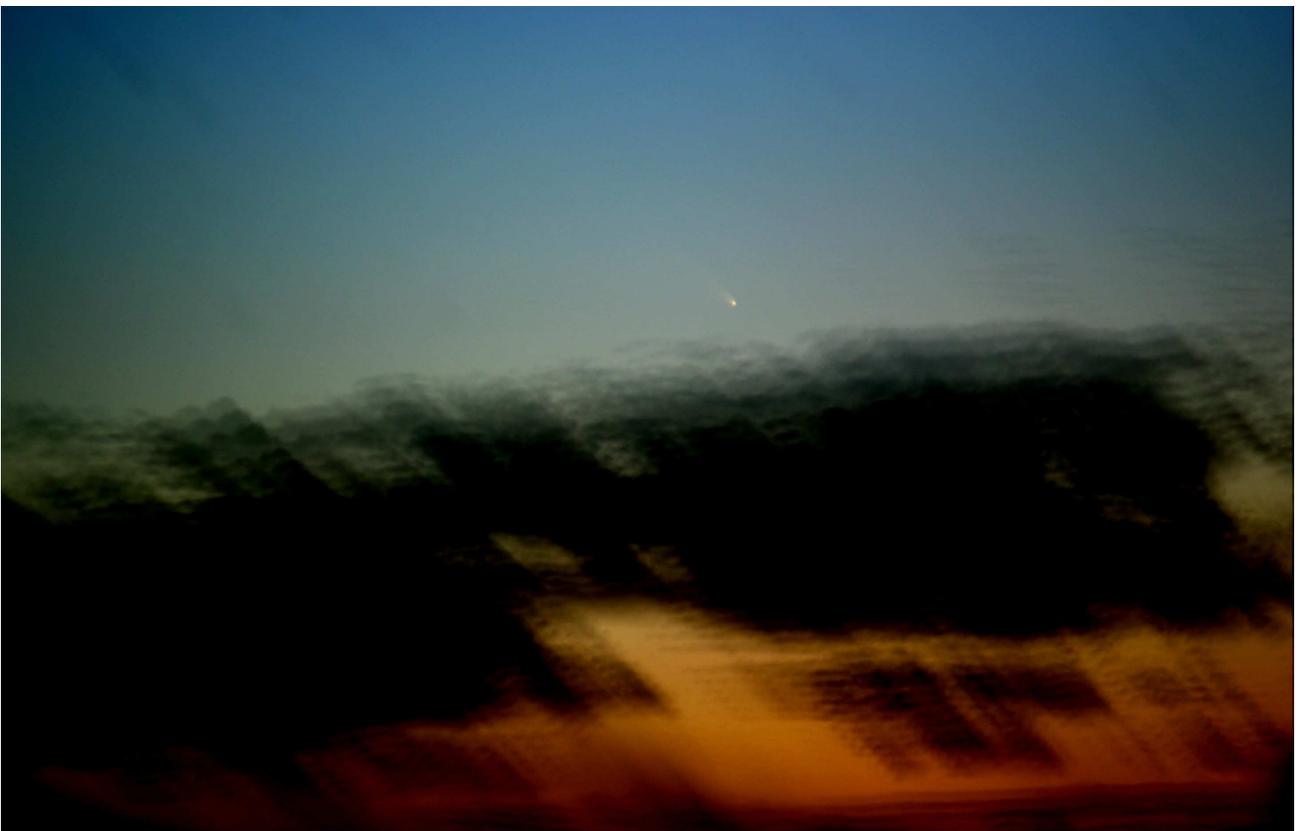

**Fig. 3 – Comet C/2011 L4 (PanStarrs), 10x1 second exposure image, taken on 2013 March 10th with Canon 10D and Tele Apo 300 mm. Note the strong condensation of the coma and the tail esteemed about 1,5° visually.**

After this shot I was able to detect easily the comet in my **7x50 binoculars**. The central condensation was marked and the tail well visible in the twilight. Stacking 10 colors images of 1 second exposure each, the comet was impressive with a tail long about **1,5°** and a coma well condensed.

**Fig.4 – 2013 March 10; single image when the comet was near the horizon level. Note the strong red color of the comet, due to the atmospheric extinction.**

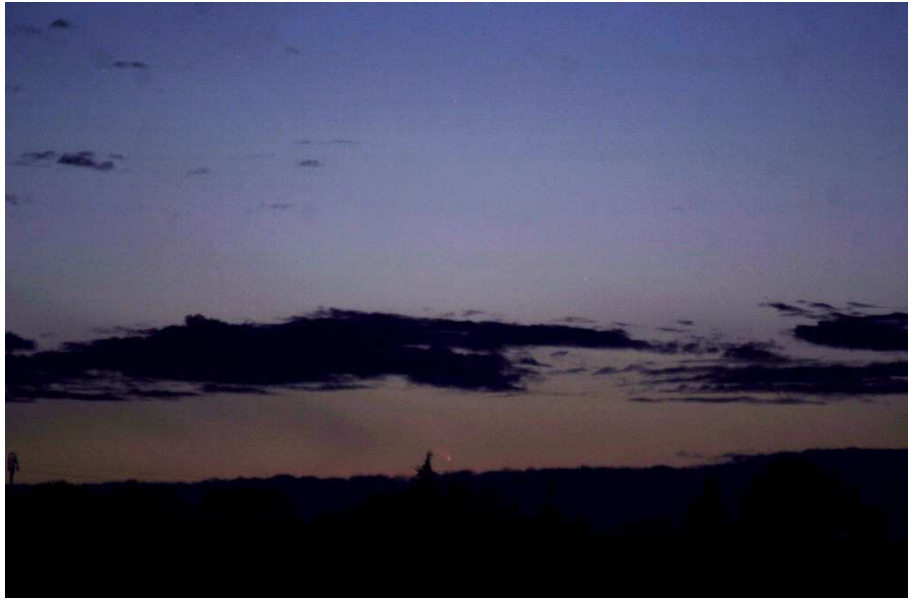

Despite the clouds above the horizon I was able to follow the comet easily visually and with my camera until to the set. Assuming an high atmospheric extinction, the apparent visual total magnitude it was esteemed -1.0 (perhaps underestimated) and the tail long about 0.5° when the comet was at about 1° above the horizon. Only **C/2006 P1** comet has had a better performance in 2007 year when reached an apparent total magnitude of -5 before to became a **"comet show"** in the southern hemisphere.

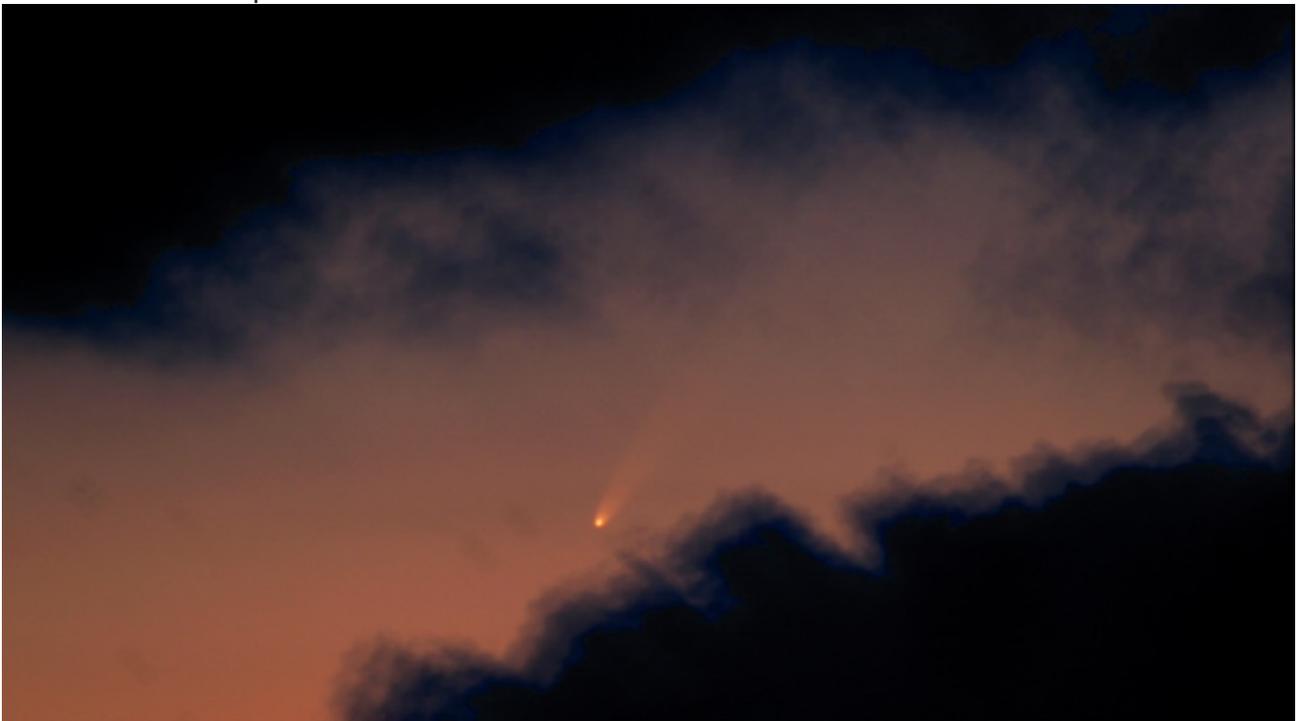

**Fig. 5 – Comet C/2006 P1 (McNaught) imaged with the same setup of the comet C/2011 L4 (PanStarrs) observation, about in the same conditions.**

By the way, in the days following the perihelion passage of the comet PanStarrs, the hope that a big show would be observed in the northern hemisphere was supported by that first observation made immediately after the perihelion passage. In addition, the flyby close to the Sun had warmed the surface of the comet with the consequent increase in activity. The goal of the following period of observation after the perihelion until to August 2013, was to determine the nucleus size, the possible period of rotation, the rate of dust production and the peculiar structures in the inner coma linked with the activity of the comet near and far from the Sun.

## 2) INSTRUMENT AND SETUP

### 2.1 CMOS and CCD CAMERA

Instruments used to observe the comet was been **CMOS Camera Canon 10D** and **CCD camera Atik 16IC**. First instrument was used with an 8 cm refractor f/5 at the direct focus of the Tele Lens Apo 70-300 mm. **Atik CCD** was used with a **25 cm Newtonian f/4.8 reflector** at the direct focus. First choice was conditioned by the position of the comet too close to the Sun. When the comet was visible in the night sky after the sunset, the CCD camera not saturate and was possible to study the coma and its inner side, until to the nucleus, in the photometric **R band**. **Canon 10D** is a discontinued **6.3 megapixel** semi-professional **digital SRL** camera. Capture RAW images in the CANON CRW file format.

| *Specification* | *Characteristics* | *Notes* |
|---|---|---|
| **Type** | Digital-Reflex | **CANON 10D** |
| **Sensor** | CMOS+IR block filter | 22,7mm x 15,1mm |
| **Pixel size** | 7,3microns x 7,3microns | 7,3 microns=7,3x10^-3 mm |
| **Maximum Resolution** | 6.3 megapixels | 3072x2048 pixels squares |
| **Shutter speed** | 1/4000 to 30 seconds | Bulb (up to 30 seconds) |
| **Asa/Iso** | 100-1600 | 3200 in extended mode |
| **Exposure metering** | TTL full aperture | Evaluative, partial, center-weighted |
| **Custom WB** | 7 presets | 2800-10000 kelvins in 100 K steps |
| **Gain** | 2,41 e-/ADU | at 400ISO |
| **RMS Noise** | 15 e- | at 400ISO |
| **Peak dark current** | 0,5 ADU x pixel/sec | at 22 °C typical |
| **Offset level** | 129adu | bias |
| **Dynamic** | 12 bit | 4095 ADU |

**Tab. 1 –** Canon 10 D CMOS camera technical characteristics.

Here below, the technical characteristics of Atik 16IC camera CCD;

| **Parameters ATIK 16IC mono** ||
|---|---|
| **Area in Pixel array** | 659 x 494 (325,546 pixels squares) |
| **Pixel size** | 7.4microns x 7.4microns |
| **Full well depth** | 40.000e- |
| **Dark current** | <1e- per second at -25°C |
| **Peak spectral response** | 500 nm |
| **Quantum efficiency** | >50% at 500 nm |
| **A-D converter** | 16 bits |
| **Readout Noise (RN)** | 7e- |
| **Anti-blooming** | yes |
| **Cooling** | yes |
| **CCD Type** | Sony ICX-424AL |
| **CCD size (*dimensions* sensitive area)** | 4,8mm x 3,7mm (dim) |

| **Parameters Telescope** || | **Parameters Filter Rc** ||
|---|---|---|---|---|
| **Aperture** | 250 mm | | **Productor** | Schuler |
| **Focal Lenght** | 1200 f/4.8 | | **Band** | Large |
| **Scale** | 1.27 arcsec/pixel | | **Lambda peak** | 5978 A |
| **Optic** | Newton | | **FWHM** | 1297 A |
| **Type** | Reflector | | | |
| **FOV (Field of View)** | 14'x11' | | | |

**Tab. 2 –** Atik 16IC CCD camera technical characteristics.

To compute the FOV (Field of view) with the two telescopes I used the following formula;

$$FOV(') = \frac{3428 \cdot \dim(mm)}{focal(mm)}, \quad scale(arc\sec/pixel) = \frac{FOV}{\dim(pixels)} \cdot 60$$

Using Canon 10D and 8 cm refractor f/5 with focal length of 400 mm, the FOV result about 194,54'x129,41' equal to 3,2degx2,2deg. The resolution of the full image is 3,8 arcsec/pixel.

## 3) FORMULAE AND ALGORITHMS

### 3.1 Differential photometry, dust production, nucleus size and images processing

To make differential photometry is necessary to captured as many photons as possible from the sources and then compare the resulting total ADU. As the number of photons that gather for a given source crucially depends on its brightness; if a source emits a steady flux, the number of photons that arrive at our detector will be settled within experimental error due to various causes, such as variation environmental conditions. If instead a source does not emit a steady flux, the number of photons that arrive will vary over time. If the change is small, for example in the order of thousandths of magnitude, then to detect it is necessary to compare the flow of variable source (comet) with the stable flux of other sources. Let us therefore refer to a simple formula which calculates the relationship between the flux of the comet in question and the total flux of other reference's sources.. The series of images, as long as possible, at least 3 or 4 hours, must to be reduced with the calibration files, dark, flat and if necessary the bias. The images are then calibrated with a program (IRIS, Astroart 3.0) to measure the number of ADU of the comet and stars in the field. When the measure for all the stars considered good according to their individual flux, the formula

$$Flux = \frac{AduComet}{\sum Adu(refs)} \text{ (Eq. 1)}$$

gives a relationship between the flux of the comet under investigation and the total flux of reference's stars. To normalize the values obtained, need estimated the median and use the relationship,

$$Fluxnormalized = \frac{Flux}{Median} \text{ (Eq. 2)}$$

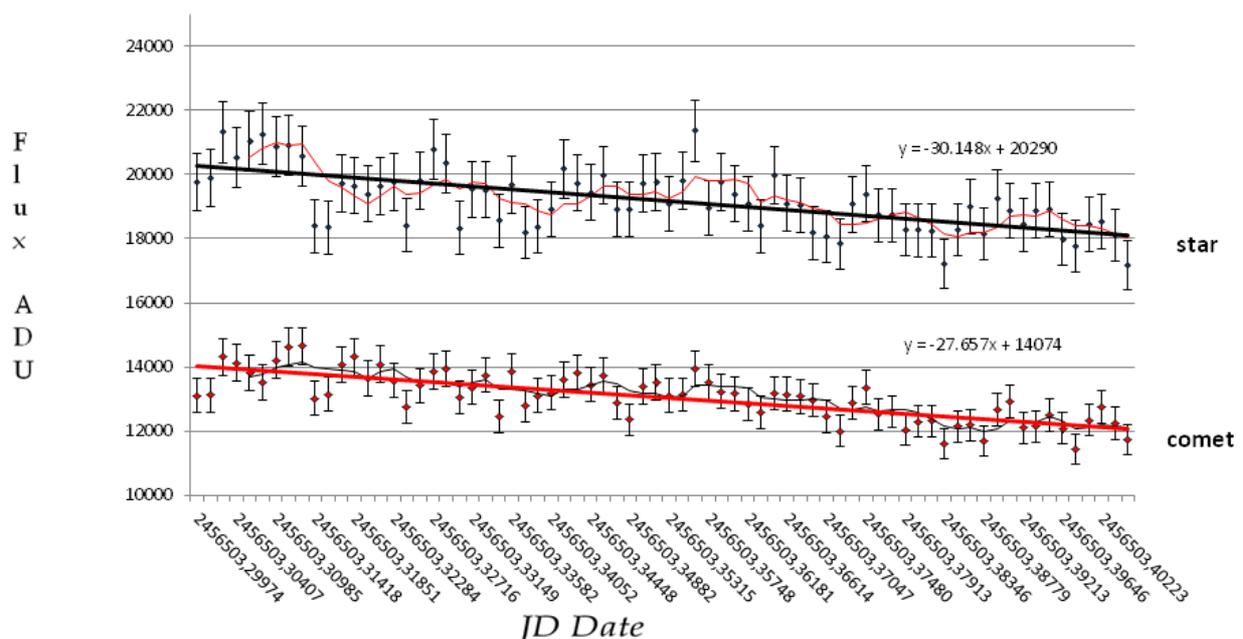

Fig. 6 – Example of comparison beetwen the comet's flux and reference's star flux.

For other details on the method see, **Toni Scarmato 2014**, http://arxiv.org/abs/1409.2693.

| Star/Catalog | mag. - USNO B1 | mag. - TYCHO 2 | mag. - SDSS |
|---|---|---|---|
| **T 2574:136:1 (Ref1)** | R1=895 – R2=8,89 | Bt=10,946 – Vt=10,629 | **R=9,340** |
| **T 2574:578:1 (Ref2)** | R1=9,77 – R2=9,69 | Bt=12,147 – Vt=10,629 | **R=10,119** |

**Tab. 3 -** Example of photometric parameters of the stars of reference derived from catalogues online using Aladin tool, after having calibrated the image and aligned on the stars.

The formula **magcomet = - 2,5 * LOG (FLUX) + magref** (where **magref** is the reference's star magnitude in **R first band USNO B1**) permit to obtain the magnitude of the comet in **R photometric band**. The measured error is the **standard deviation** at **1 sigma** precision. The **S/N (Signal/Noise)** ratio is directly linked with the uncertainties on the magnitude; to convert the value **S/N** in magnitude's error I used the formula,

$$\sigma(m) = +/- 2,5 \cdot [\log(S+N) - \log(S)] \text{ (Eq. 3)}$$

Assuming that there are other variables that produce uncertainties in magnitude, I compute $\sigma(m)$ as standard deviation on all the observations. **Af(rho)** is the dust production rate of the comet that we observe on the comet's coma **(A'Hearn 1984)**. To compute the value was used the following formula:

$$Af(rho) = (2,467 \cdot 10^{19}) \frac{R_s^2 \cdot d}{ap} (\log Fluxcomet - \log FluxSun) \text{ (Eq. 4)}$$

that with simple transformation becomes,

$$Af(rho) = (2,467 \cdot 10^{19}) \frac{R_s^2 \cdot d}{ap} 10^{0,4(M_S - M_C)} \text{ (Eq. 5)}$$

where **Rs** is the Sun-Comet distance in A.U., **d** the Earth-Comet distance in A.U., **ap** is the used aperture in arcsec, **Ms=-27.09** the magnitude in **R** band of the Sun and **Mc** the measured magnitude in R band of the comet. The method to estimate the nucleus size of a comet with images at lower resolution from the ground, was discussed in detail in **Toni Scarmato 2014**, http://arxiv.org/abs/1405.3112. Measure the ADU for the nucleus and using the stars in the FOV of the images and his R magnitude, compute the radius of the comet. Using the following model of the luminosity, **(Lamy et al. 2009);**

$$L(\rho) = [k_n \delta(\rho) + coma] \otimes PSF \text{ (Eq. 6)}$$

with a comet's model,

$$Model = [nucleus + coma] \otimes PSF \text{ (Eq. 7)}$$

where **PSF** denotes the **Point Spread Function** of the telescope and $\otimes$ represents the **Convolution** operator. To resolve the nucleus, in the sense that it is possible to measure the real spatial size, we can use the following equation;

$$nucleus = k_n \delta(\rho) \text{ (Eq. 8)}$$

where **δ** is the Dirac's delta function, *ρ* is the radial distance from the center, and $k_n$ is a scaling factor. The apparent magnitude *Ma* in *R* band of a comet can to be linked to its physical properties by the equation,

$$p \cdot \Phi(\alpha) \cdot R_n^2 = 2{,}238 \cdot 10^{22} \cdot R_S^2 \cdot d^2 \cdot 10^{0{,}4(M_S - H)} \quad \text{(Eq. 9)}$$

where **p** is the geometric albedo, *Φ(α)* the phase function, *Rn* is the radius of the comet in meters. With simple transformations, we can obtain the comet radius in meters;

$$R_n = \frac{1{,}496 \cdot 10^{11}}{\sqrt{p}} \cdot 10^{0{,}2(M_S - H)} \quad \text{(Eq. 10)}$$

where

$$H = M_a - 5 \cdot \log(R_s) \cdot d - \alpha \cdot \beta \quad \text{(Eq. 11)}$$

is the **_absolute magnitude_** of the nucleus in **R band**, and

$$\alpha \cdot \beta = -2{,}5 \cdot \log[\Phi(\alpha)] \quad \text{(Eq. 12)}$$

$$\Phi(\alpha) = 10^{-\frac{\alpha \cdot \beta}{2{,}5}} = 10^{-0{,}4 \cdot \alpha \cdot \beta} \quad \text{(Eq. 12 bis)}$$

where *Φ(α)* is the phase function, *α* is the phase angle, **Ma** is the apparent magnitude of the nucleus measured from the observations in **R** band and **β** the phase coefficient. **Rs=1 A.U., β=0.04 mag/deg** and the albedo **p=4%** was assumed as standard values as reported in the literature, also if, other values was computed for single comets. In fact, it is possible to find also **β=0.035 mag/deg and different p** when were made independent measures.

To process the images have been applied some powerful algorithms, based on filters and mathematical computations. The complete procedure follow the steps as labeled here below: Bicubic Interpolation, Convolution+PSF, RWM and MCM filters. In the section 4.5 there are other details on the procedure.

## 4) ANALISYS OF THE RESULTS

### 4.1 First observations results

On **2016 March 3th** evening, good weather allow me to observe the comet using CMOS Canon 10D camera.

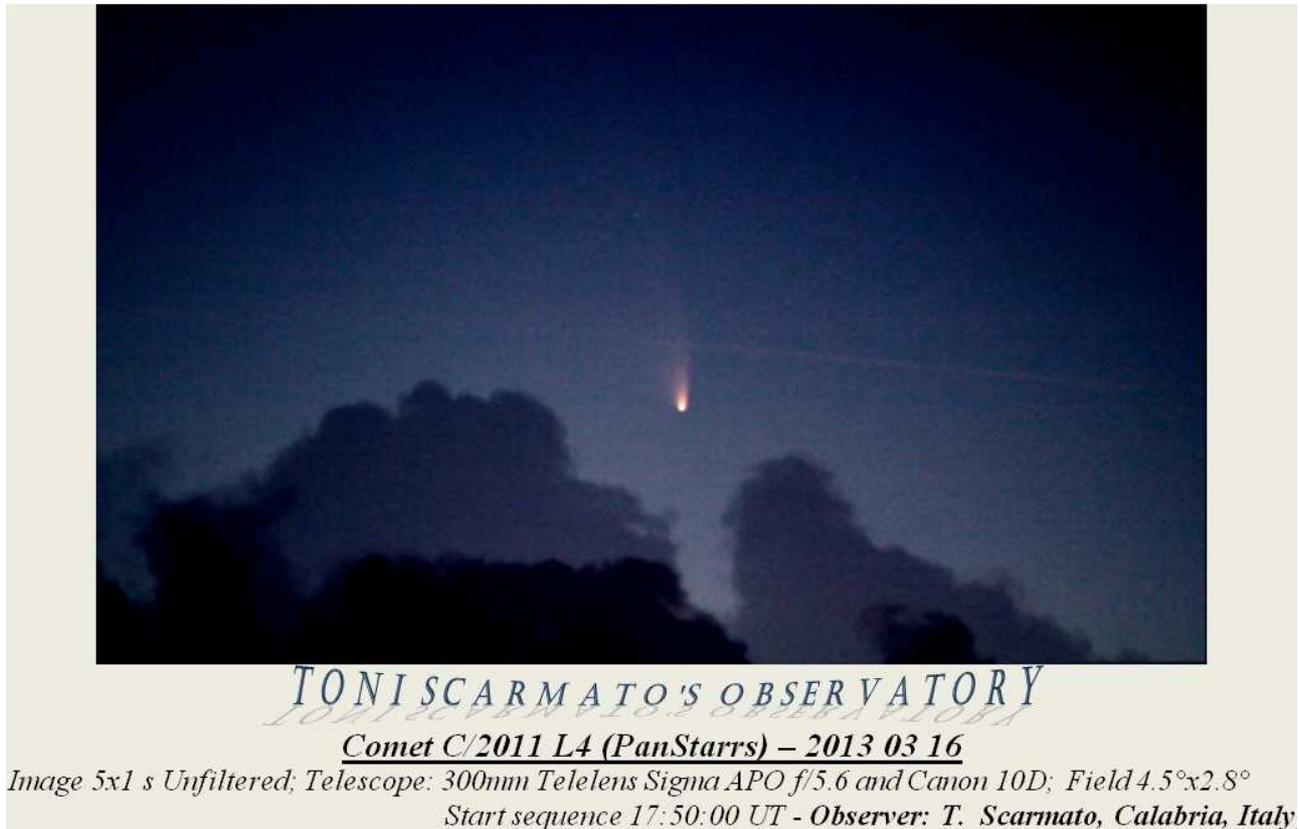

Fig.7 – Comet C/2011 L4 imaged in the twilight on 2013 March 16th.

Is interesting note that, despite the twilight, are clearly visible the strong condensation and a fan shaped tail. An analysis of the tail and inner coma show interesting details. (**see following Fig.7bis, Fig. 8**)

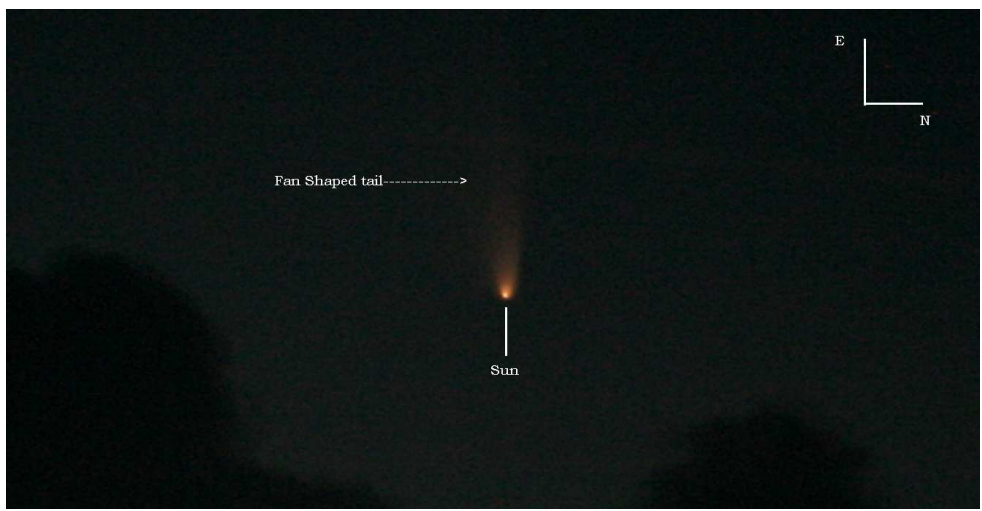

The strong central condensation and the clear tail are signs of strong activity of the comet, probably due to the heating of the all nucleus surface due to the close passage to the Sun at perihelion. That suggest me to try to measure **afrho** parameter in the following days. Also, the enhancement of the inner coma in all three bands R,G,B, show the fan shaped tail clear in R band but not a clear central pixel. That suggest that the comet is producing a large quantity of dust and that the nucleus is probably small.

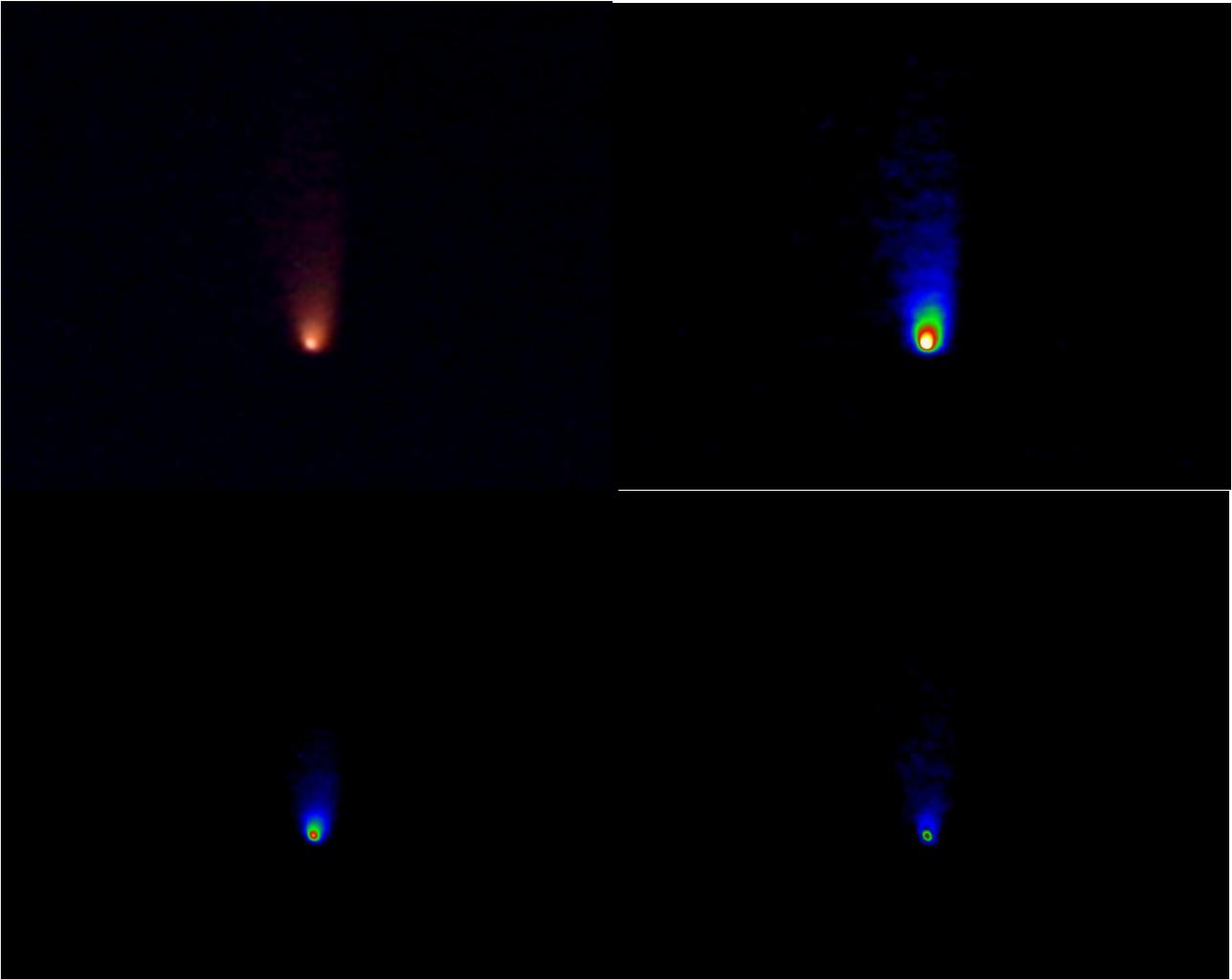

**Fig. 8 – Top left: Original RGB color image, Canon 10D and 300 mm Telelens Apo f/5.6, resolution 3.8 arcsec/pixel, taken on 2016 March 16; spatial resolution at comet distance is about 3320km/pixel; top right: image in R band 580-700 nm; bottom left: image in G band 490-560 nm; bottom right: image in B band 430-490 nm. Est at top, North at right.**

Also if the separation of the three channels R,G,B, for an image taken with the CMOS sensor as that of Canon 10D, is not much selective, it is possible note the very different shape of the comet in the relative bands. Unfortunately, in the field, there are not stars to do photometry because the comet was in the twilight and the exposure was of only 1 second, so I waited one good evening to measure the magnitude of the comet. Surely, an analysis of the inner coma show very interesting results. How it is possible to see in the following image (**Fig. 9)**, there are four regions with different shape and brightness. Central area containing the nucleus show an apparent uniform brightness; the region at about **1670 km** show an interesting clear separation zone between the inner coma and the outer asymmetric region at about **3120 km** from the nucleus. Also, that region show a boundary that look like a separation between the two zones from **3120 km** to the outer coma at about **10400 km**.

*Comet C/2011 L4 (PanStarrs)*

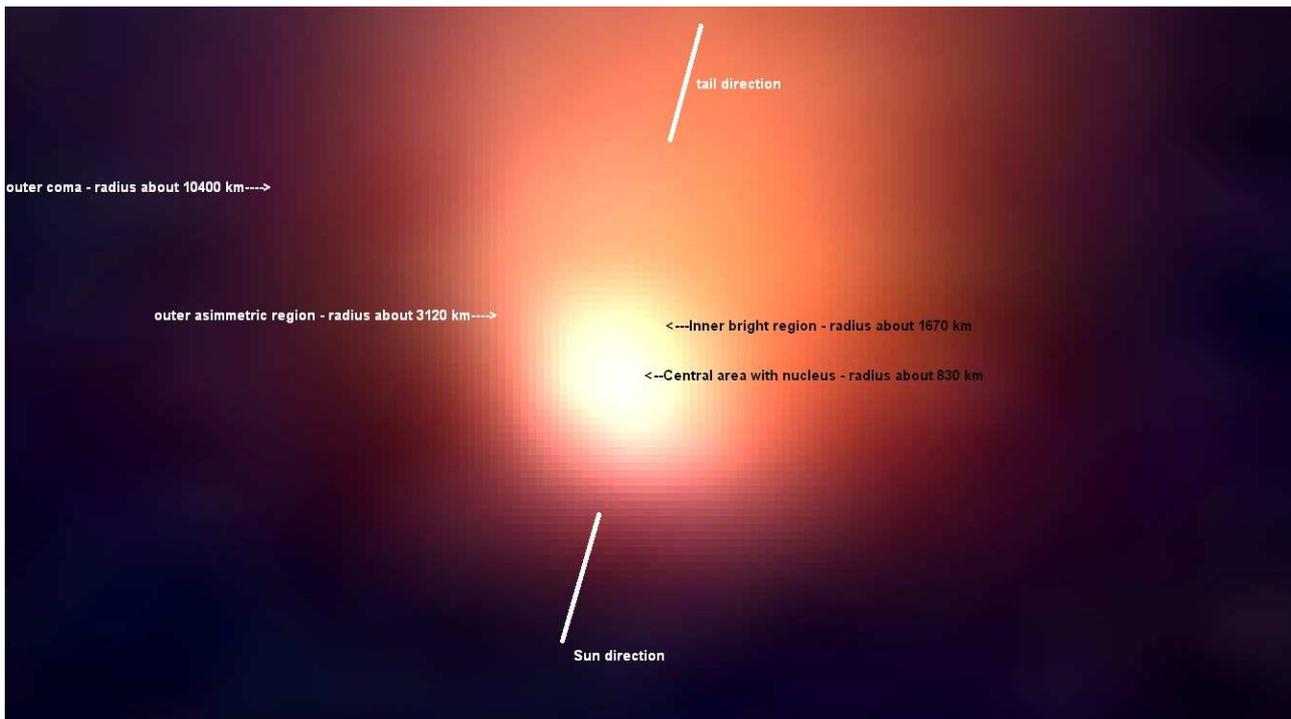

**Fig. 9 – Resampled image with Bicubic Interpolation of the color RGB image, Canon 10D and 300 mm Telelens Apo f/5.6, original resolution 3.8 arcsec/pixel, taken on 2016 March 16; spatial resampled resolution at comet distance is about 208 km/pixel. Est at top, North at right.**

Also if the resolution of those images is low, this details have suggested me to perform as possible as precise the photometric analysis of the comet in R,G and B band. Further analysis of the central area in the resampled image using the **Bicubic Interpolation** (see **Toni Scarmato 2014**, http://arxiv.org/abs/1409.2693), show that the pixel containing the nucleus is not clearly separated from the surrounding area. The pixelization of the area with the Bicubic Interpolation works on the 16 pixels around the position for which we want the value and permit to obtain the finer structure of the image. On our image I used 4x4 resampling to divide one pixel in 16 sub-pixels to identify the finer **GRADIENT OF BRIGHTNESS** in the coma. As said previously, that suggest that the nucleus can to be small. By the way with that image is not possible to apply the complete process to measure the nucleus size. Need to have CCD image with higher resolution and higher dynamic. We can only search possible structures in the inner coma. Because I suspect also the possibility that the central area is saturated, due to the fact that there are 8 pixel with the same brightness, to confirm this results I have compared that image with another image of another comet captured with the same camera Canon 10D. The observation of the comet **C/2013 R1** was made on **2013 Dcember 11th**. The stacked image of **5x30 sec exposure** at the direct focus of **25 cm f/4.8 Newtonian** show an intense emission surrounding the nucleus at **Est Sud-Est direction**. As it is possible see, RGB image show different emissions that, the separation of the channels, confirm to be "**dust**" the structure pointed as "**Dust Jets**". (see **Figures. 10 and 11**).

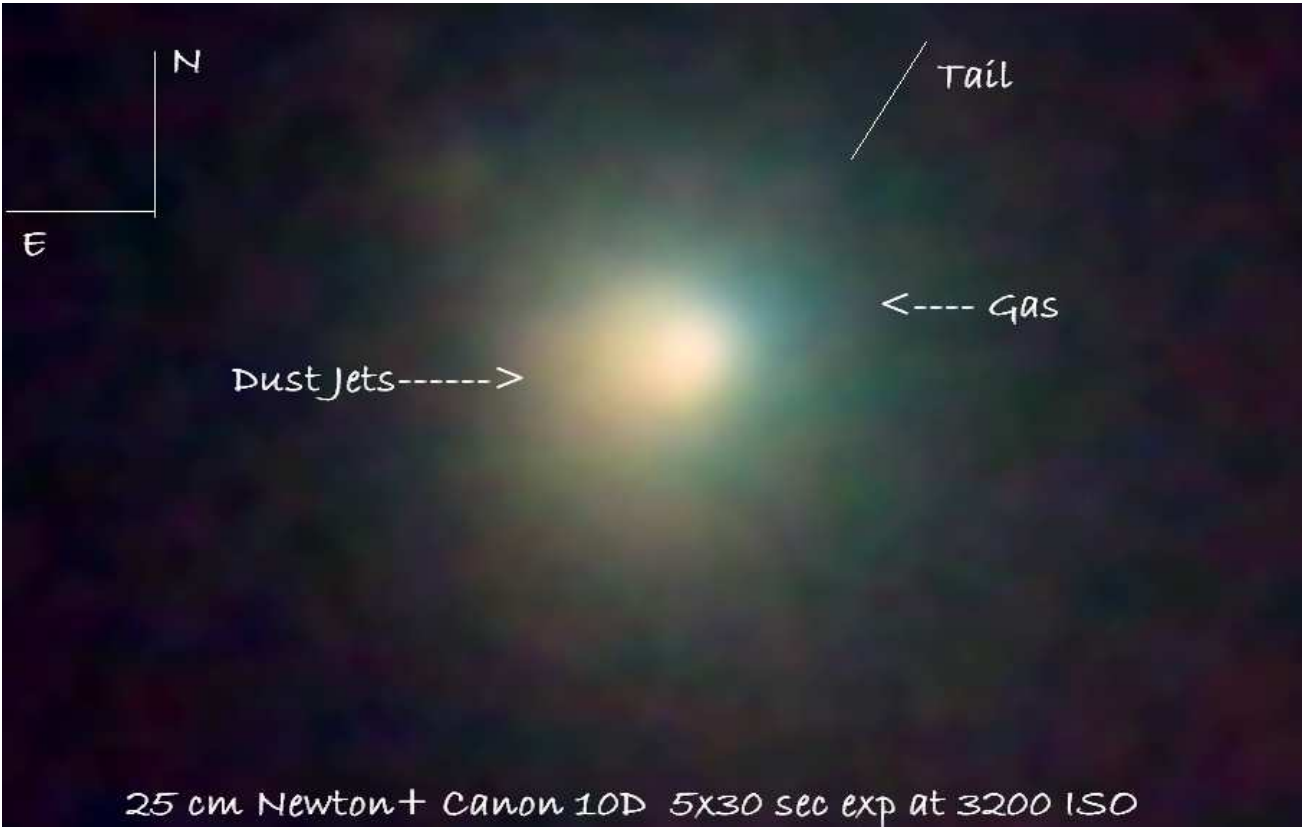

**Fig. 10 – Original image of comet C/2013 R1 taken on 2013 December 11 with instrument in the caption.   That comet will be analized in another article in preparation.**

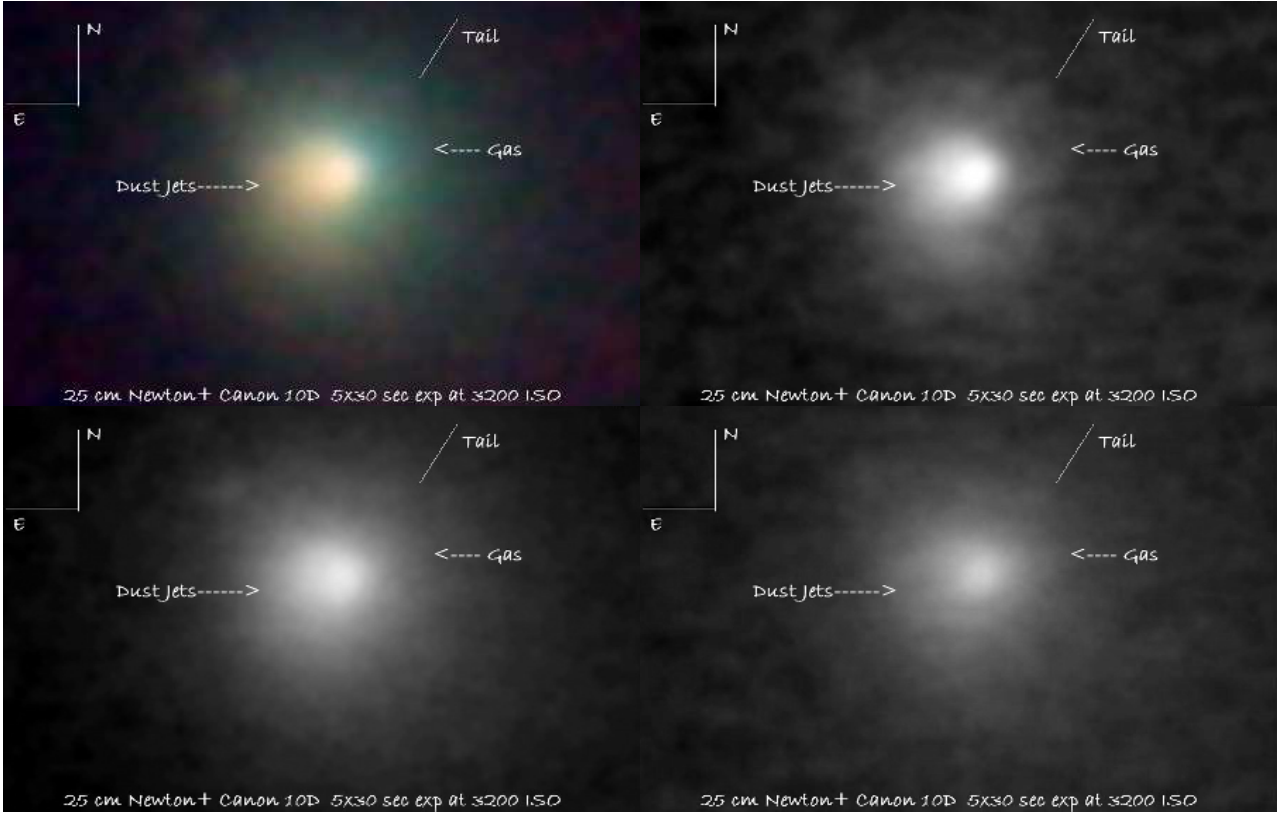

**Fig. 11 – Top left: original RGB image; Top right: R channel; Bottom left: G channel; Bottom right: B channel**

Also, CCD images in Rc band **(see Fig. 12)** show the same structure around the nucleus.

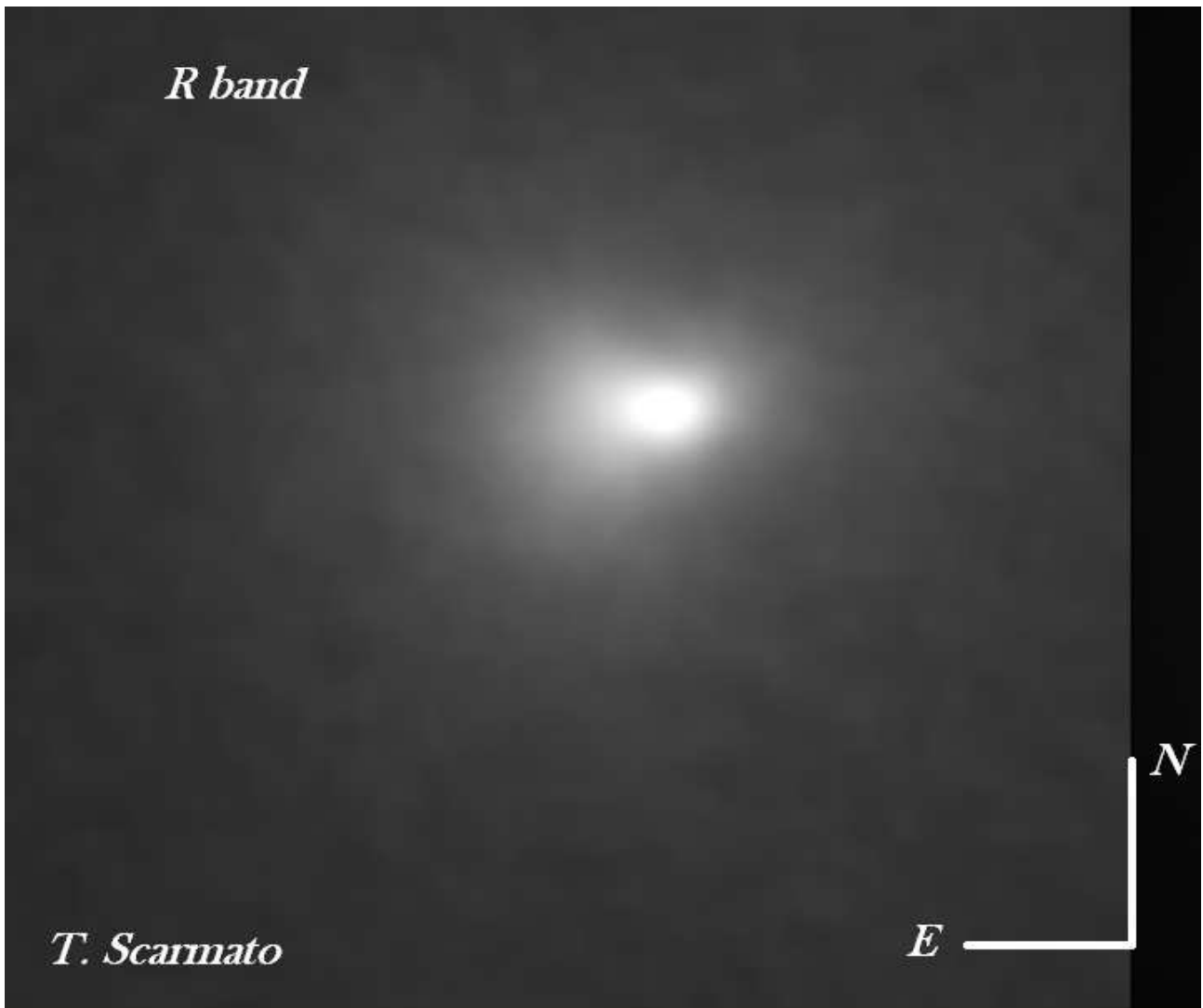

**Fig. 12 – CCD images of comet C/2013 R1 in R band.**

Those observations are strong evidences about the fact that RGB images with CMOS detectors, also if are not much selective, can give good indications about the emissions in the cometary comae. This comparison is also a strong evidence that the coma of **C/2011 L4** is in large part composited by **"dust"** emitting in the range **580-700 nm**, with a probable very low "**gas**" component in the range **430-560 nm**.

### 4.2 Differential Photometry on RGB images

In the evening on **2013 March 19$^{th}$ (see Fig. 13 at right)** the comet was well positioned in the sky but, also if the great dust tail was clear visible, I don't been able to image star in the field to do photometric measurement. The fan shaped dust tail appear large and curved toward **Sud**, no sign of ions tail also if the exposure time and the brightness of the sky can interfere with the probable faint emission. The good night to do photometry was on **2013 March 23th** with the comet well positioned in the **N-O sky** at about **30°** above the horizon with a sky sufficiently dark to perform higher exposure time. In the following image we can see the evolution of the dust tail, no sign of ions tail and some good photometric star in the field. Time exposure of **120 seconds** allowed me to obtain a good **S/N (signal to noise)** both for comet and star used for the photometry (see Fig. 14).

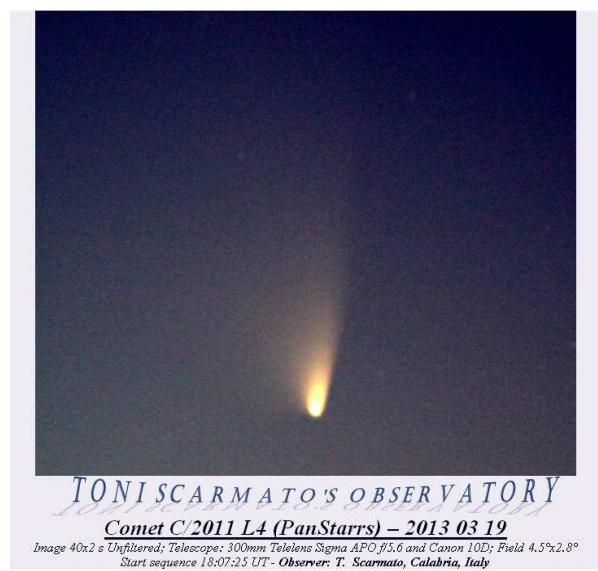

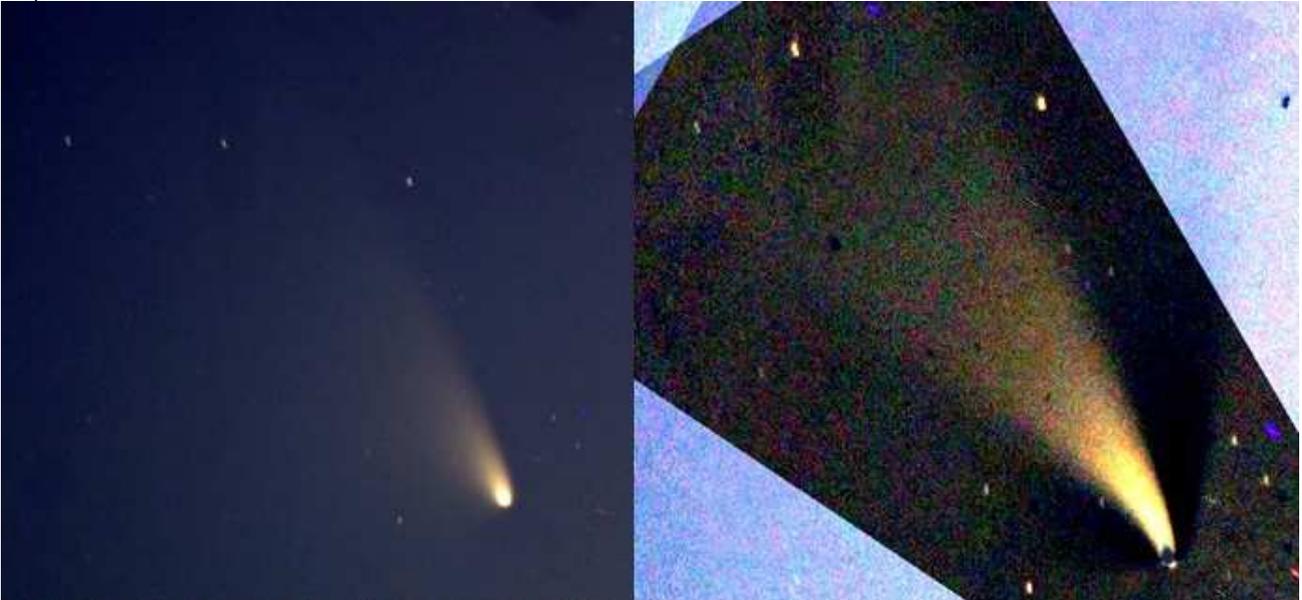

**Fig. 14 – Comet C/2011 L4 (PanStarrs) taken on 2013 March 23th, with Canon 10D and 300 mm Telelens Sigma Apo f/5.6; left: original RGB 5x120 sec image; right Larson-Sekanina filter applied. North is at the top, Est at left.**

Looking at the spectral response of Canon 10D here below,

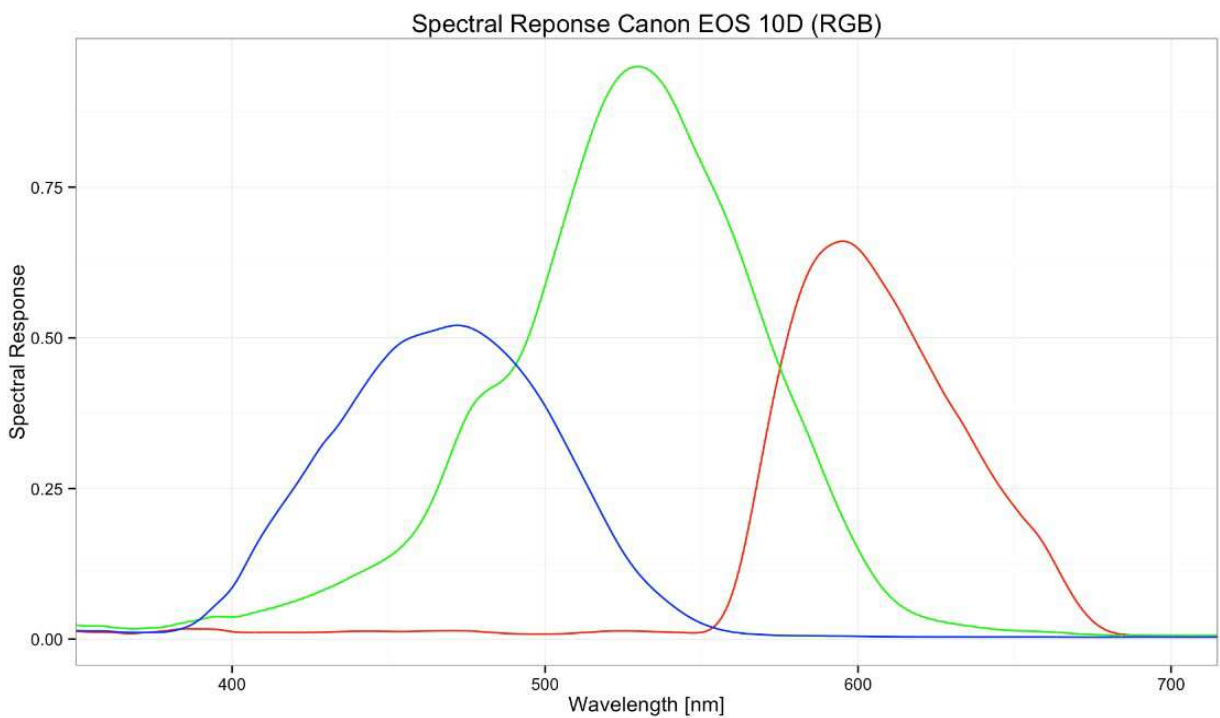

**Fig. 15 – Note the complete absence of response at lambda >680 nm due to the IR block filter on the camera's CMOS.**

and at the following panel (figure 16) showing the comet and the relative RGB channels, and assuming that **G channel** is equivalent to the **V band (490-560 nm)**, we can extract preliminary qualitative information about the emission in the coma and tail.

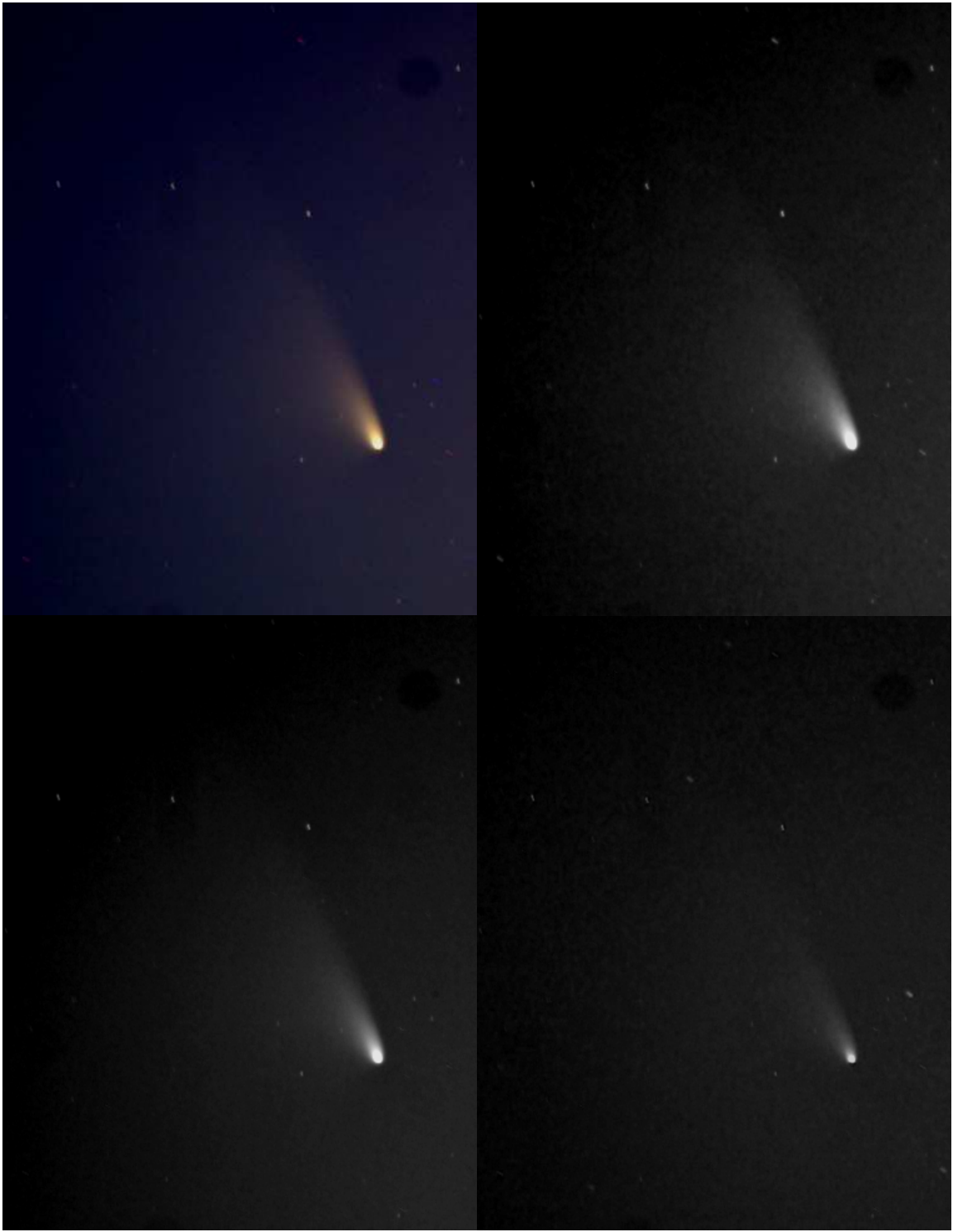

**Fig. 16 – Comet C/2011 L4 (PanStarrs) taken on 2013 March 23th, with Canon 10D and 300 mm Telelens Sigma Apo f/5.6; Top left: original RGB 5x120 sec image; Top right R channel; Bottom left G channel, bottom right B channel. North is at the top, Est at left.**

Apparently, the situation of the comet look like changed since **2013 March 16th** observation. The R channel image show apparently the same brightness of the G channel. B channel instead, continue to be the band with low emission. Then, after that qualitative and not so much detailed analisys, we pass to do measurements of the magnitude. To do this I have identiefied a good star in the field.

**HD3204 is a G spectral type** star that was the best in the field to compare with the comet flux. In the following table the characteristics of the star extracted using **Aladin** and several photometric catalogs online.

| *HD 3204 - G Spectral Type Star – AR 003525 Decl. +250613* ||||||||| 
| Bands: B(400-500 nm) – V(500-600 nm) – R(600-750 nm) – I(>750 nm) ||||||||| 
| Catalog/Mag(band) | *B* | *V* | *R(1/2)* | *I* | *B-V* | *B-R* | *V-R* | Notes |
|---|---|---|---|---|---|---|---|---|
| *HIP (Hipparcos)* | 9,576+/-0.022 | 8,150+/-0.013 | | | +1,426 | | | |
| *HIP2 (Hipparcos)* | | 8,1611+/-0,0018 | | | | | | |
| *USNOB1* | | | 7,370/7,290 | | | | | |
| *UCAC4* | 9,465+/-0,012 | 8,868+/-0.004 | 8,794+/-0,008 | | +0,597 | +0,671 | +0,074 | |
| *SDSS-DR9* | 8,773+/-0,001 | | 7,637+/-0,001 | 7,401+/-0,001 | | +1,136 | | |
| *THYCO-2* | 9,577+/-0,024 | 8,154+/-0,013 | | | +1,423 | | | |

**Tab. 4 – HD 3204 color index.**

The following table show the photometry of the comet in all the bands RGB. For the **B and V bands** I used **HIP/Thyco-2** magnitude and for **R band USNO-B1 R1** magnitude, for the comparison star, good for the precision of the measurements.

| *Comet C/2011 L4 (PanStarrs) RGB photometry on 2013 March 23th image* |||||||||
| Star HD 3204 – G Spectral Type – Flux (ADU)-mag (band) |||||||||
| *Obj/Mag* | *Flux R* | *Flux V* | *Flux B* | *MagR* | *MagV* | *MagB* | *B-V* | *B-R* | *V-R* |
|---|---|---|---|---|---|---|---|---|---|
| *Comet* | 31799 | 19922 | 8105 | 3,161+/-0,530 | 4,204+/-0,650 | 6,223+/-0,702 | +2,019 | +3,062 | +1,043 |
| *Star* | 659 | 526 | 368 | 7,370+/-0,010 | 8,150+/-0,013 | 9,576+/-0,022 | +1,200 | +2,000 | +0,800 |

**Tab. 5 – C/2011 L4 differential photometry and color index.**

As it is possible notice, the errors are very large due to the fact that the **S/N ratio** is low. By the way, the fluxes and the relative magnitudes of the comet show that we have the most emission in R band with a good flux also in V band and low flux in B band. Considering the study in "**Comet classification with new methods for gas and dust spectroscopy**" Laura E. Langland-Shula' ,Graeme H. Smith/ Icarus 213 (2011), the bands observed are relative at two molecules of gas, **C2 and NH2**, respectively at **455-475 nm** and **568-578 nm**. For the dust is not possible to determine the size of the grains, because that is a qualitative more than quantitative analysis. Surely, it is possible and then important measure the dust rate production using *Af(rho)* parameter.

## 4.3 Af(rho) and dust production

The evening on **2016 March 23th** and **2013 March 30th** the comet show really to be a dust-rich comet and confirm the data obtained on March 16th. The main indicator of cometary activity is the **dust/gas ratio**. If we want simplify the description, the dust–gas ratio of a comet depending as the comet turn around the Sun, loses the easily sublimated ices near the surface, and develops a dusty crust.   The dust production parameter Af(rho), first computed in A'Hearn et al. (1984), is useful for narrowband aperture photometry that should be carefully selected to only contain dust continuum, and not contaminate by gas emission bands.  Using the band R selected from RGB image, we can assume that we work in the range 580-700 nm, and looking at the spectral response of CMOS of the Canon camera we can also assume that in that range there is no great contamination of gas emitting in the V band.  That assumption is not strictly quantitative, but surely is a good indication of the dust production and the results look like in good agreement with other more precise data in the literature.

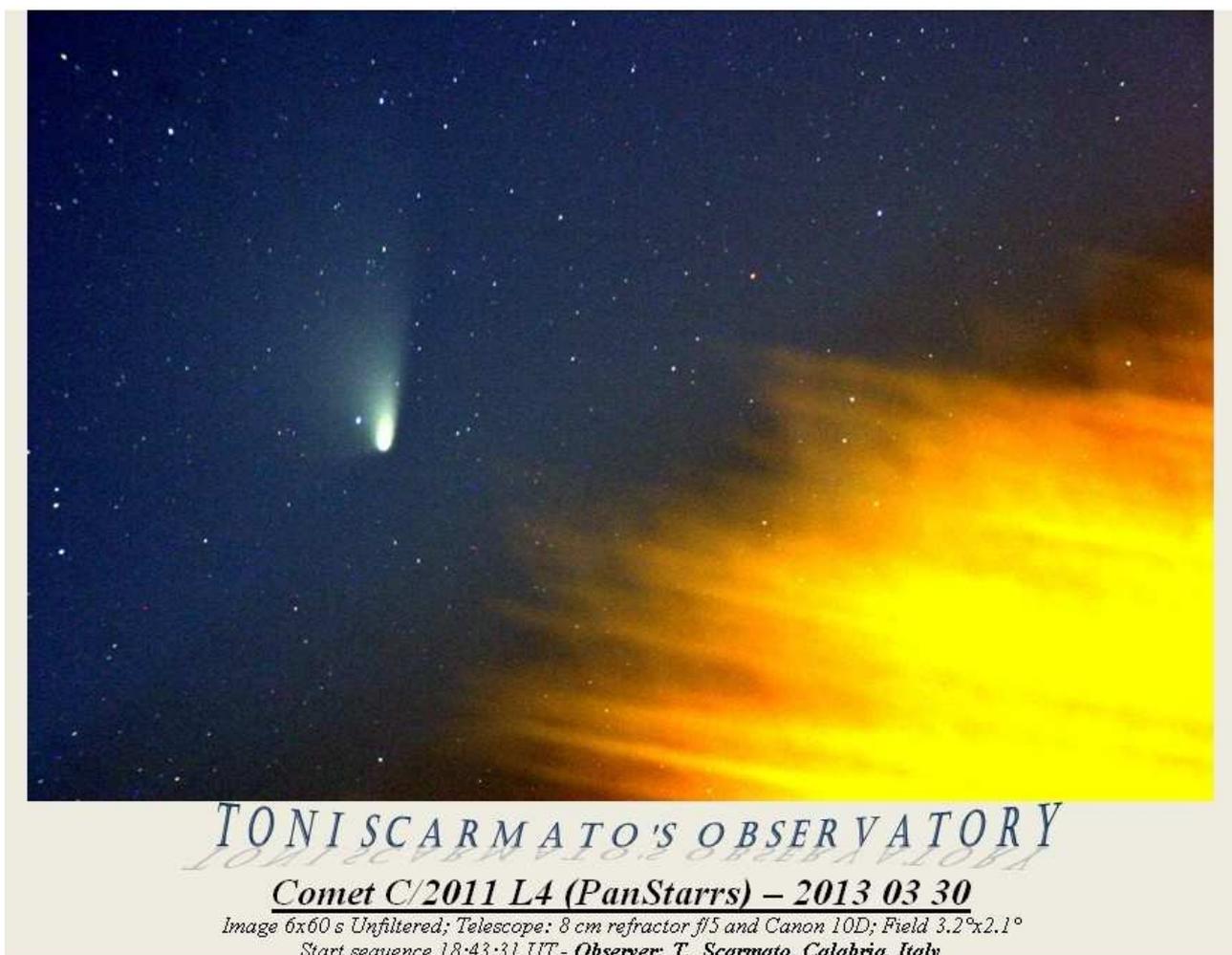

**Fig. 17 – Comet C/2011 L4 (PanStarrs) imaged on 2013 March 30th with an 8 cm refractor and Canon 10D.**

Particularly interesting is the image of the figure 17 above.  Taken on **2013 March 30** with a good dark sky, show a very large fan-shaped dust tail.  The evolution of that tail will be confirmed to be mainly dust because has formed a very long anti-tail in May-June months when the Earth crossed the orbit-plan of the comet.  **Af(rho)** value was about **2700 m** and the measurement show that the comet was strongly active in March month.

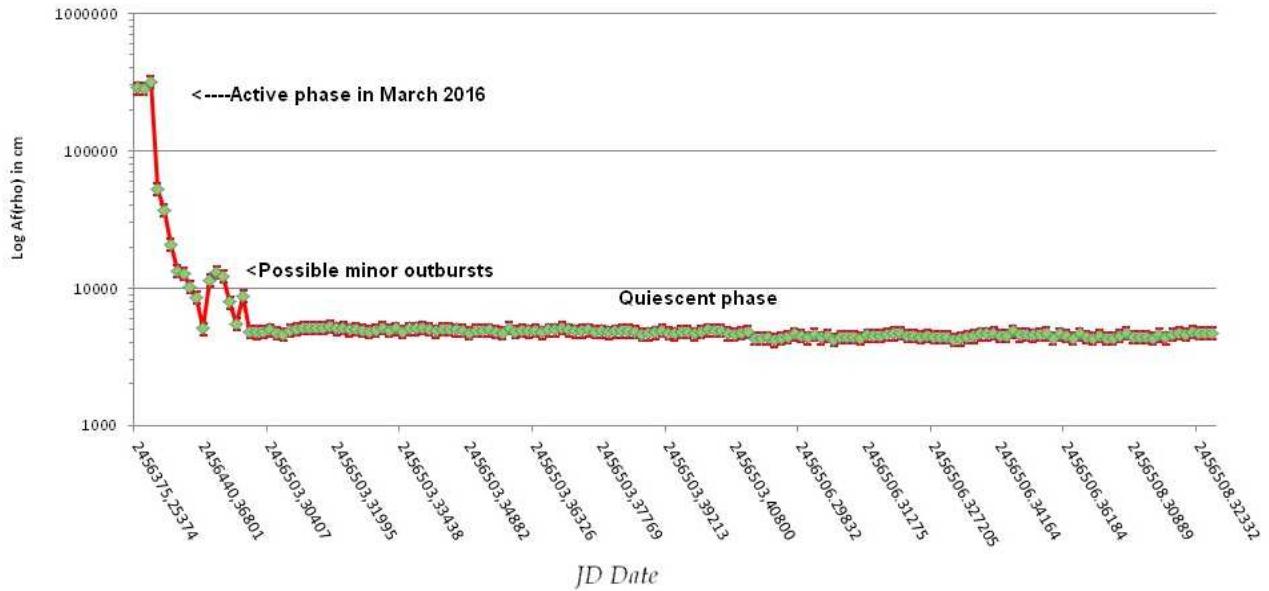

**Fig. 18 – Af(rho) of C/2011 L4 measured after perihelion.**

**The data are corrected for phase's angle,** the trend of **Af(rho)** value after perihelion increase until the end of March, to decrease after that time. The data of the **"quiescent phase" (July-August),** have been used to compute the period of rotation and the nucleus size.
In the graph below (figure19) are visible two curves; green line plot the regular theoretical Light Curve of the comet in visual range, red line plot the Light Curve measured in R band corrected for the phase function. The general formula for the Light Curve, contain the apparent magnitude *m* for specific coma's aperture, the absolute magnitude $m_0$ for *D* and r equal to 1 A.U. and the **n parameter**;

$$m = m_0 + 5\log D + 2{,}5n\log r .$$

The best fit of the red curve is the following formula:

$$m_R = 5{,}16 + 5 \cdot \log D + 15{,}17\log r .$$

The regular formula, green curve, is fitted by the equation,

$$m_V = 5{,}5 + 5 \cdot \log D + 10{,}0\log r .$$

Interesting to note that the absolute magnitude is almost the same in the two range of emission (red and visible), instead the coefficient **2,5n** is much different. The **n parameter** is linked with the **"nature"** of the comet structure. For the red curve we have **n=6,1** that, in the past theory of the classification of the comets, indicate a **"dusty"** nature of the comet structure.

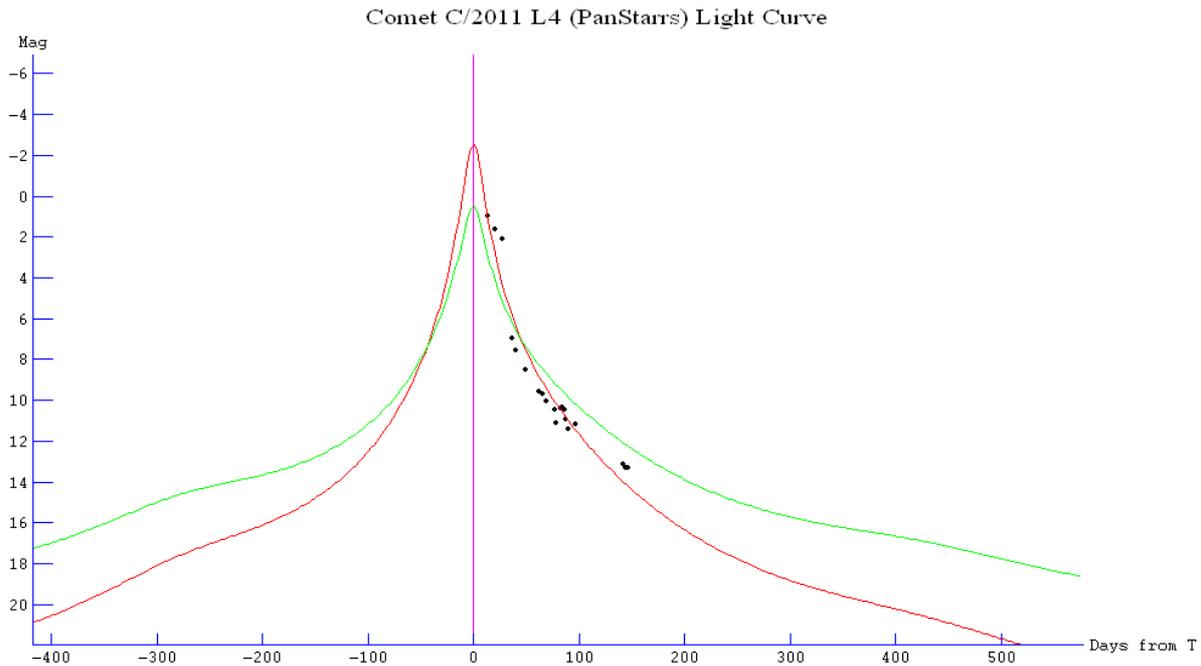

**Fig. 19 – Light Curve of comet C/2011 L4.** Red line plot the observed R band trend, gren line the visual theoretical trend.

To compute the dust rate production I used the light curve in figure 18. The best fit found is given by the following power law;

$$Af(rho) = 43230 \cdot r^{-0.517} (cm)$$

with ***r distance from the Sun***. At perihelion ***r=q=0,3015433 A.U.*** we have the following rate of dust production any second;

$$Q(dust)_{sec} = 43230 \cdot (0,3015433)^{-0.517} = 80345 \cdot 10^0 (kg/s)$$

that is a considerable amount of dust in one day,

$$Q(dust)_{day} = 6,942 10^9 (kg/day)$$

In the quiescent phase (July-August observations), instead, considering a linear relation between **Af(rho) and r** we have an average value at second;

$$Q(dust)_{sec} = 4350 \cdot 10^0 (kg/s)$$

and lost in one day;

$$Q(dust)_{day} = 3,76 \cdot 10^8 (kg/day)$$

This value is a relevant amount of dust lost at about 3 A.U. from the Sun also if of the order of 10 times less than in March-April, period of very strong activity. In the table 6 I have summarized the Af(rho) and Qdust values for all the observations after perihelion.

| C/2011 L4 (PanStarrs) photometry | | | | | | | |
|---|---|---|---|---|---|---|---|
| *Date* | *Mag* | *Af(rho) cm* | *α °* | *D (A.U.)* | *r (A.U.)* | *Ap " (arcsec)* | *Qdust (kg/day)* |
| 20130323 | 1,001+/-0,161 | 267319+/-6683 | 54,0 | 1,2020 | 0,4860 | 152 | 5,42x10^9 |
| 20130330 | 1,651+/-0,191 | 266870+/-6671 | 51,8 | 1,2590 | 0,6400 | 152 | 4,70x10^9 |
| 20130406 | 2,100+/-0,178 | 298286+/-7457 | 49,3 | 1,3170 | 0,7930 | 144 | 4,21x10^9 |
| 20130416 | 6,987+/-0,029 | 49165+/-734 | 45,8 | 1,4000 | 1,0020 | 17 | 3,74x10^9 |
| 20130419 | 7.561+/-0,017 | 34249+/-513 | 44,7 | 1,4830 | 1,0630 | 17 | 3,62x10^9 |
| 20130428 | 8,530+/-0.025 | 19432+/-291 | 41,7 | 1,5020 | 1,2380 | 17 | 3,34x10^9 |
| 20130511 | 9,568+/-0.031 | 12433+/-186 | 37,7 | 1,6210 | 1,4770 | 15 | 1,07x10^9 |
| 20130514 | 9,709+/-0.027 | 11939+/-179 | 36,8 | 1,6500 | 1,5310 | 15 | 1,03x10^9 |
| 20130517 | 10,040+/-0.032 | 9588+/-143 | 35,9 | 1,6810 | 1,5830 | 15 | 8,28x10^8 |
| 20130525 | 10,480+/-0.018 | 7931+/-119 | 33,7 | 1,7670 | 1,7200 | 15 | 6,85x10^8 |
| 20130527 | 11,099+/-0.019 | 4728+/-70 | 33,2 | 1,7900 | 1,7540 | 15 | 4,08x10^8 |
| 20130531 | 10,427+/-0.024 | 10583+/-159 | 33,2 | 1,8370 | 1,8200 | 14 | 9,14x10^8 |
| 20130602 | 10,331+/-0.019 | 12140+/-182 | 31,7 | 1,8610 | 1,8530 | 14 | 1,05x10^9 |
| 20130604 | 10,451+/-0.027 | 11413+/-171 | 31,2 | 1,8860 | 1,8850 | 14 | 9,86x10^8 |
| 20130605 | 10,949+/-0.016 | 7378+/-110 | 31,0 | 1,8980 | 1,9020 | 14 | 6,37x10^8 |
| 20130607 | 11,404+/-0.021 | 5085+/-76 | 30,5 | 1,9240 | 1,9340 | 14 | 4,39x10^8 |
| 20130614 | 11,175+/-0.025 | 8105+/-121 | 28,9 | 2,0180 | 2,0450 | 10 | 7,00x10^8 |
| 20130729 | 13,135+/-0.018 | 4584+/-76 | 21,3 | 2,7690 | 2,7090 | 7 | 3,96x10^8 |
| 20130801 | 13,294+/-0.023 | 4168+/-85 | 20,9 | 2,8250 | 2,7500 | 7 | 3,60x10^8 |
| 20130803 | 13,307+/-0.022 | 4298+/-89 | 20,7 | 2,8630 | 2,7780 | 7 | 3,70x10^8 |

**Tab. 6 -** R band photometry corrected for phase angle and phase coefficient 0,04 mag/deg. D Earth-Comet distance, r Sun-Comet distance.

### 4.4 Period of rotation and brightness variations

**Converting the flux in magnitude** we note the trend of the brightness with a descending phase, possible minor outbursts and an **"apparent" quiescent phase on three days,** with an interesting beaviour, to the end of July and first days of August.

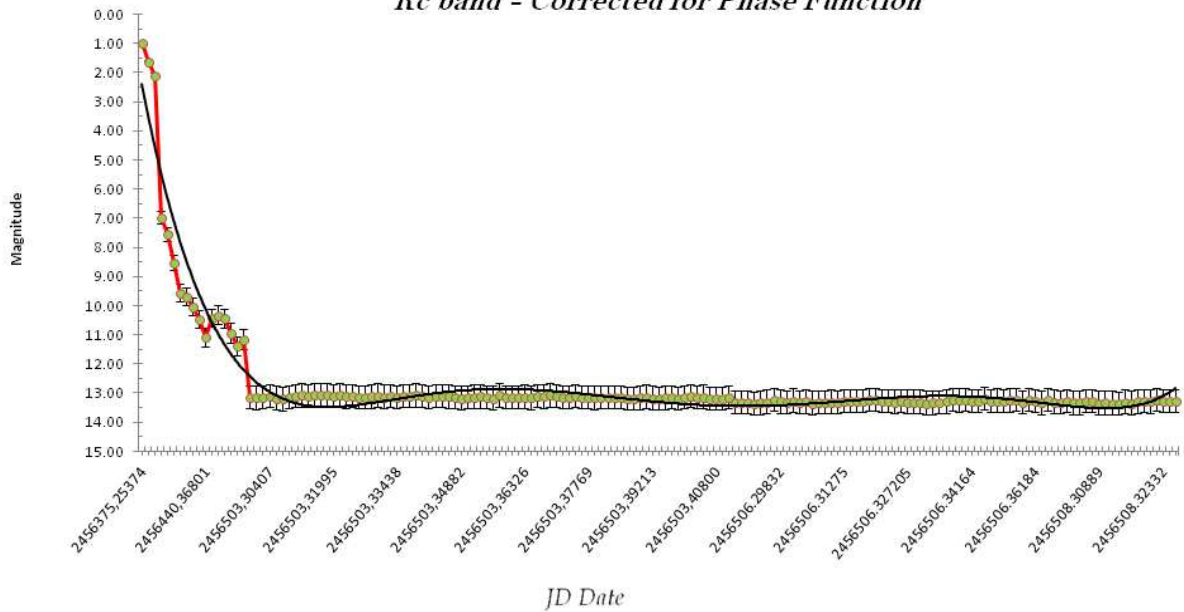

**Fig. 20 – Light Curve I R band of comet C/2011 L4. Note the trend during the quiescent phase.**

**Detailed analysis of the "apparent" oscillation in the trend,** during the "**apparent**" quiescent phase, using an algorithm able to extract possible periods in a time-series data, show that the **periodic variation of the brightness** in the time was real with a clear period. I observed the comet on 3 days obtaining 3 time-series data. Differential photometry and normalization of the flux have provided good results. The flux normalized show a variation in the time strongly supported from the comparison between the comet flux and the stars flux because there is the same trend during the time-series data. The observations have been made in three night from **2013 July 29th** to **2013 August 3th** with **25 cm Newton f/4.8 telescope and CCD camera Atik 16Ic with Rc filter**. The technical characteristics of the instruments are in the table 2 above. For any night the images have been calibrated with darks and flats and the comparison's star was in the FOV.
The procedure is that describe above in 3.1, and in **Toni Scarmato 2014**, http://arxiv.org/abs/1409.2693, based on the differential photometry. As said before (**see also example in Fig. 6**), to verify if both comet and star's flux have had the same trend during all the observation, I plotted the normalized data for any single observation's night. On **2013 July 29th** the comet was at 66° of altitude in the Bootis constellation at the start of the observation, 19:11:37 U.T. (Universal Time). The distance from the **Sun was 2,7090 A.U.** and from the **Earth 2,7690 A.U.** I follow the comet just to 21:51:41 U.T. so for two hours and 40 minutes. I obtained 75 images in Rc band with exposure of 120 sec to have a good S/N. Stars used **Usno-B1 1424-0289473** and **Usno-B1 1424-0289550** at the end of the observation (last 4 images). As it is possible see in the figure 21 the two fluxes are almost over imposed with a small deviations so we can assume a good precision in the differential photometry.

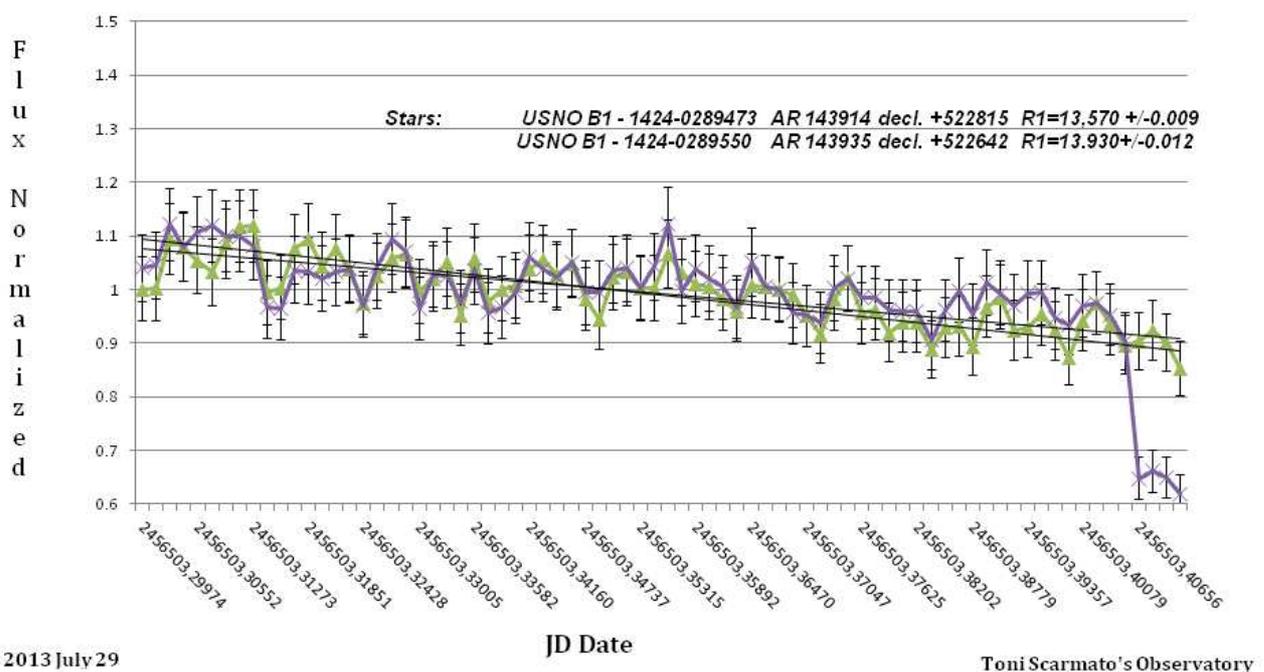

Fig. 21 – 2013 July 29th observation. Comet and comparison star fluxes normalized and overimposed.

The measured magnitude of the comet corrected for the phase function is *m1=13,135+/-0,018* in Rc band, *phase=21,3°*, *Af(rho)=4584+/-76 cm* for an aperture of 7 pixels equal to 9 arcsec equivalent to about *18000 km* at the comet distance.

On **2013 August 1th** the observation's conditions don't change much with the comet at 63° of altitude, *phase=20.9°*, started at 18:55:01 U.T. ended at 20:57:40 U.T. so about two hours of observation. I measured 45 tracked images in Rc band, exposure 120 sec. Star used, in the FOV, **Usno-B1 1411-0245369**. Distance of the comet from the **Sun 2,7500 A.U**. and from the **Earth 2,8250 A.U.** Magnitude of the comet in Rc band **corrected for phase function**, *m1=13,294+/-*

***0,023***, **_Af(rho)=4168+/-85 cm_** for 7 pixels of aperture. The comet seem much active yet, also if not as the 2013 March's month. The dust production continue to be the main emission and considering the distance from the Sun, the cause can be the complete or almost complete sublimation of the surface's ice at the close perihelion passage that have, much probably, uncovered all crust of dust. As we will see in the following analysis in this paragraph, the comet result a fast rotator so can be all the surface emitting dust.

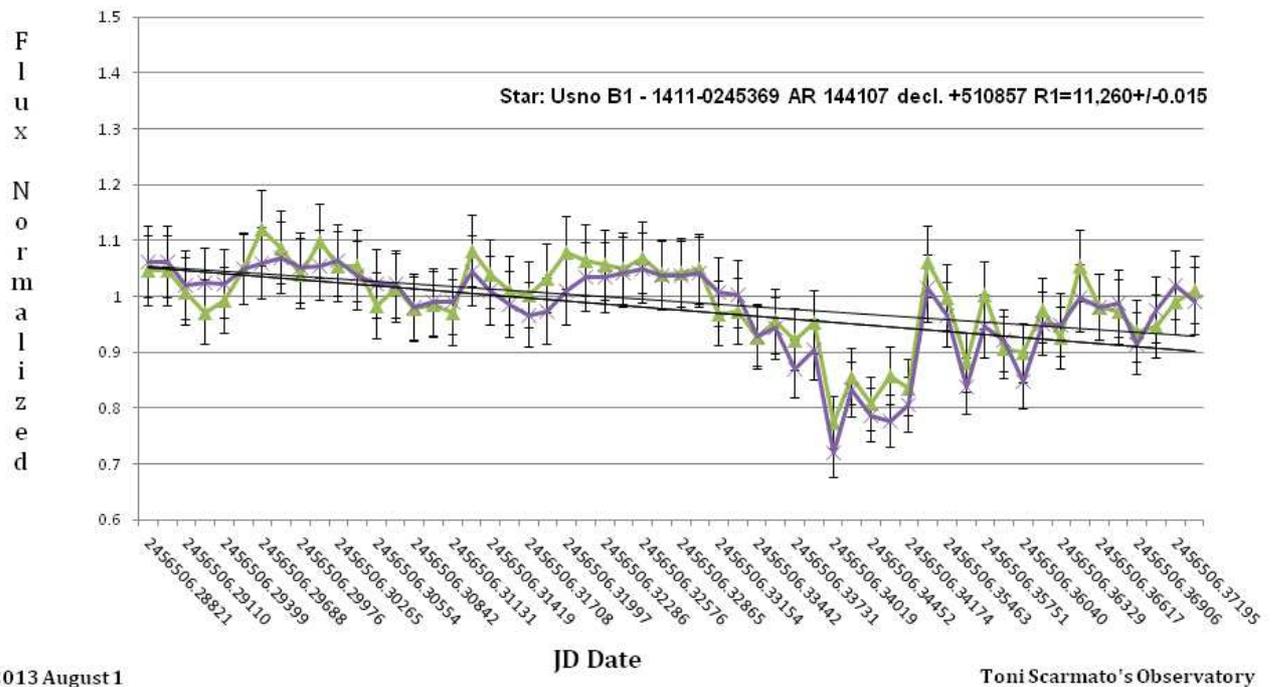

**Fig. 22 - 2013 August 1th observation. Comet and comparison star fluxes normalized and overimposed.**

Another good night was on **2013 August 3th**. The comet was observed for about 30 min with the comet at 60° of altitude**, _phase=20,7°_**, from 19:22:43 U.T. to 19:49:44 U.T in Rc band 120 sec exposure time. Star in the FOV **Usno-B1 1402-0269693**, **Sun-Comet distance 2,7780 A.U. and Earth-Comet 2,8630 A.U.** Comet magnitude **corrected for phase function _m1=13,308+/-0,022_** and **_Af(rho)=4298+/-95 cm_**. Aperture 7 pixels equal to 9 arcsec. Measurements confirms the **steady state** of the comet also if the Af(rho) value put in evidence a good activity in the dust production, with a tendency to a slow decrease.

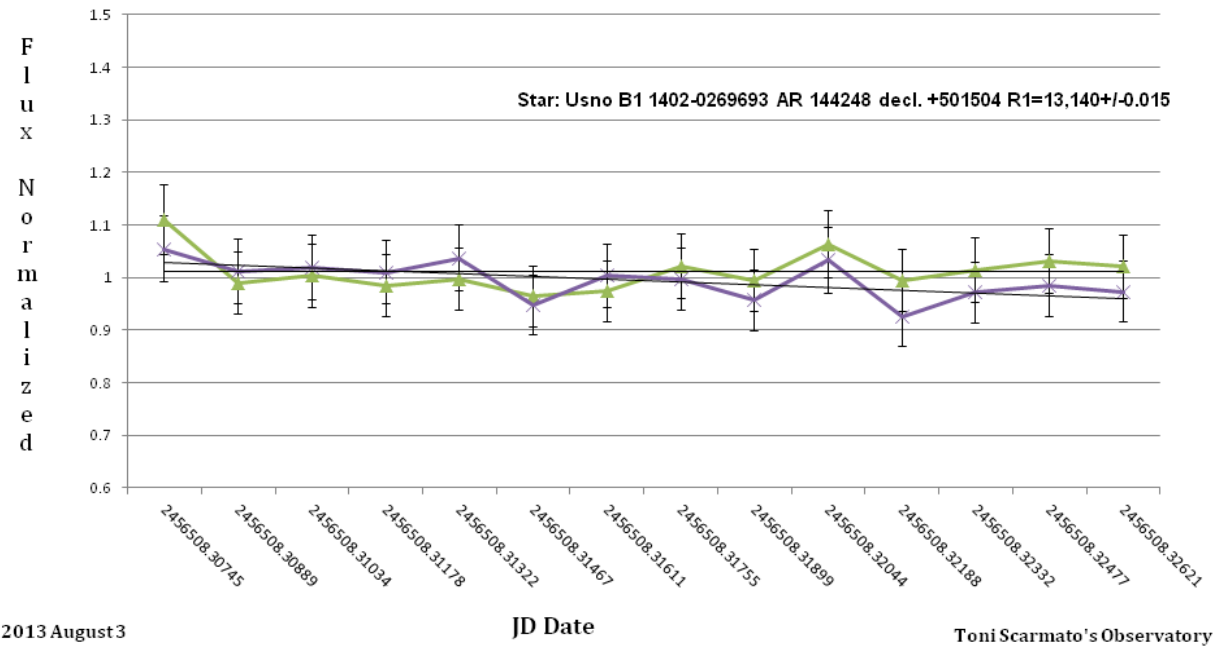

**Fig. 23 - 2013 August 3th observation. Comet and comparison star fluxes normalized and overimposed.**

Plotting all the data relative to the differential photometry of the comet with respect the stars of comparison and normalizing the flux, I obtained an interesting trend that clearly show a variation of the brightness of the comet in the time.

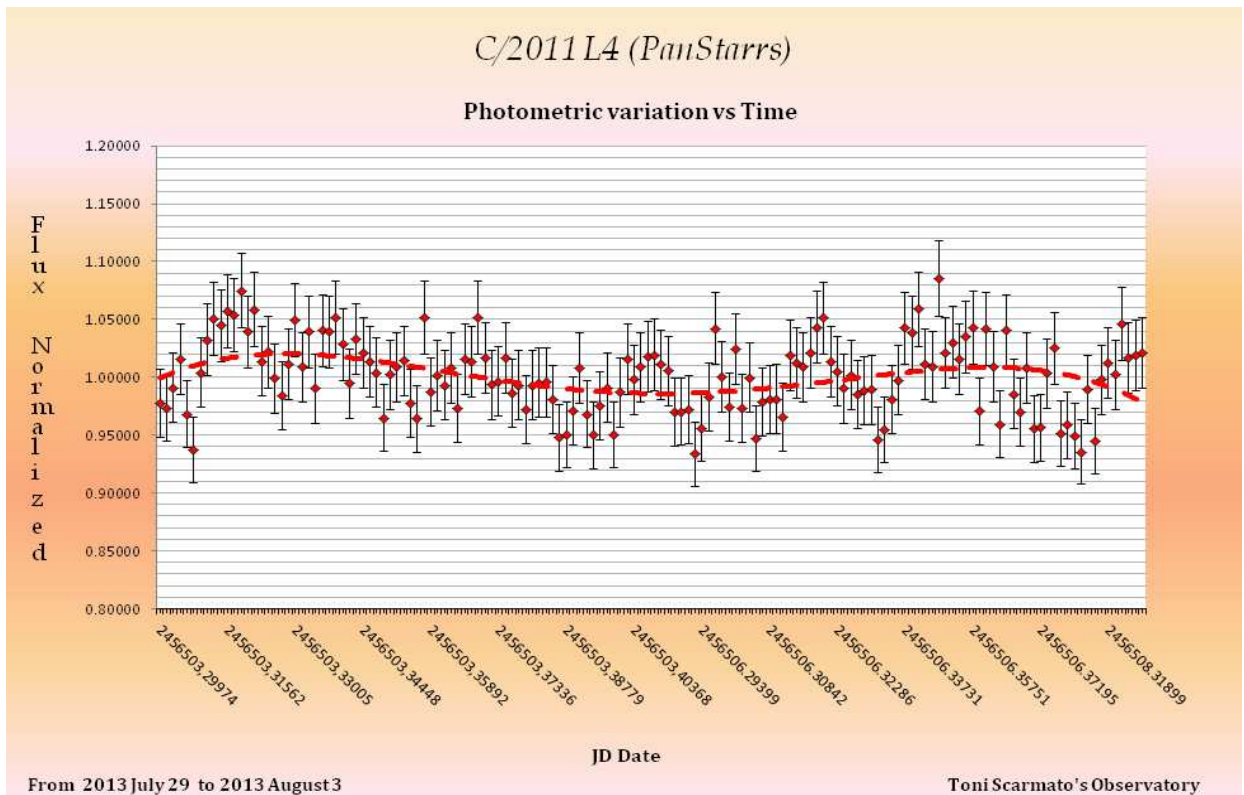

**Fig. 24 – Normalized flux show photometric variations on three days from 2013 July 29th to 2013 August 3th.**

Using **Plavchan PDM algorithm**, I detect the sinusoidal variation in the light curve. That periodic time-series shapes can be linked with the rotation of the nucleus. The analysis show a clear rank with **period of 4.8999 hours** as showed in the periodogram below.

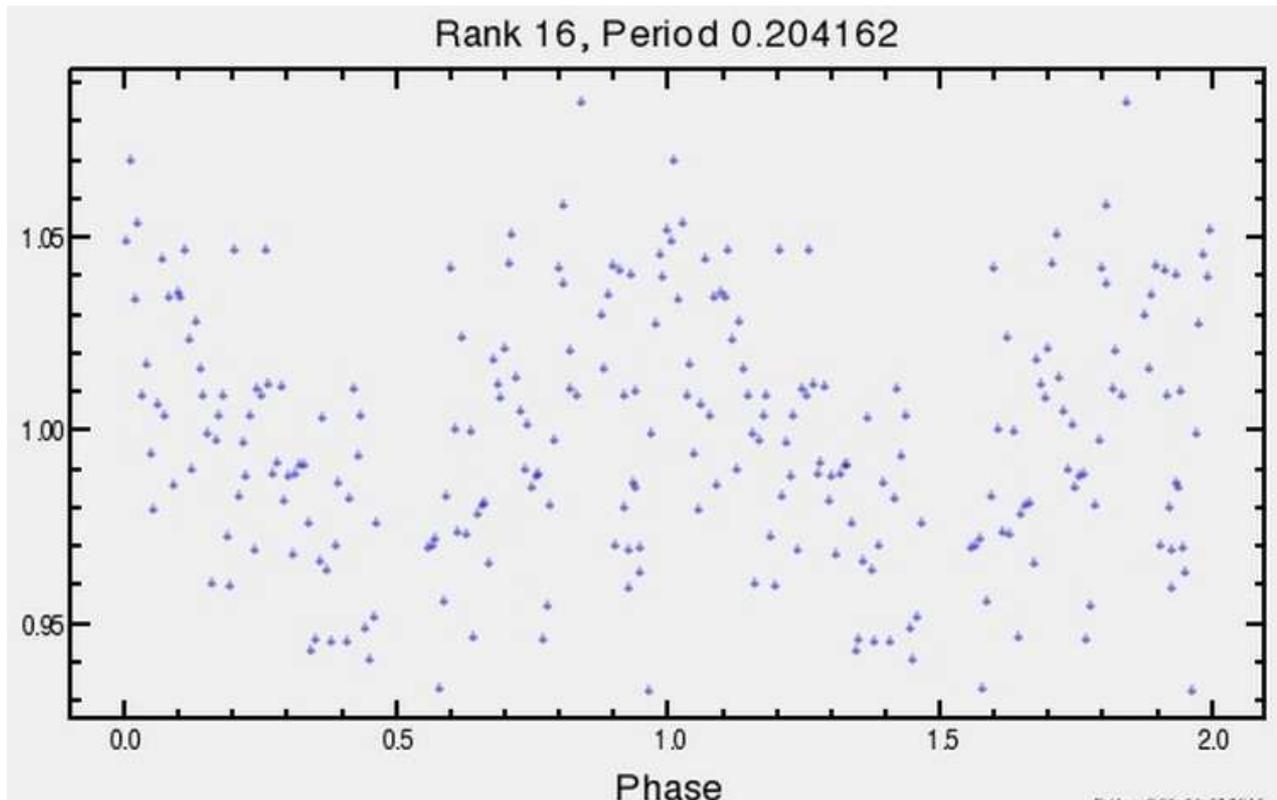

**Fig. 25 – Periodogram using Plavchan algorithm show a sinosoidal beaviour with period of 4,8999 hours.**

### 4.5 Nucleus size and coma profile

The study of the size of cometary nuclei and of their physical properties, is important to understand the formation and evolution of the Solar System. Up to this time, the determination of the nucleus size of a comet, was made only with Hubble telescope images. The camera used with HST has a resolution of 0.04 arcsec/pixels. Because the cometary nuclei are much smaller than spatial resolution, one needs to use the algorithms to separate the contribution of the nucleus from the coma, to the total brightness coma + nucleus. **(Lamy et al. 2009)**. I have developed one method to measure the nucleus of a comet also with images at lower resolution (see Table 2). I applied this method to C/2012 S1 (ISON) comet (**Toni Scarmato 2014**, http://arxiv.org/abs/1405.3112) obtaining good results in agreement with professionals result (**P.Lamy, I. Thot, H.A. Weaver; "Hubble Space Telescope Observations of the Nucleus of Comet C/2012 S1 (ISON)"; The Astrophysical Journal Letters 794(1):L9 · September 2014).** In this work I did a more detailed study to analyze the profile of the inner coma around the nucleus. Results are interesting and seem in agreement with a coma profile not homogenous along different directions. On the other hand, recent observations of comets with spacecraft, (Deep Impact, Rosetta) show that the coma around the nucleus is not homogenous due to the rotation of the nucleus and "jets" emitting material in the space (ice and dust). Look at the image of 103/P comet taken from Deep Impact. It is clear that the jets coming out form the nucleus producing a gradient of the coma brightness distribution.

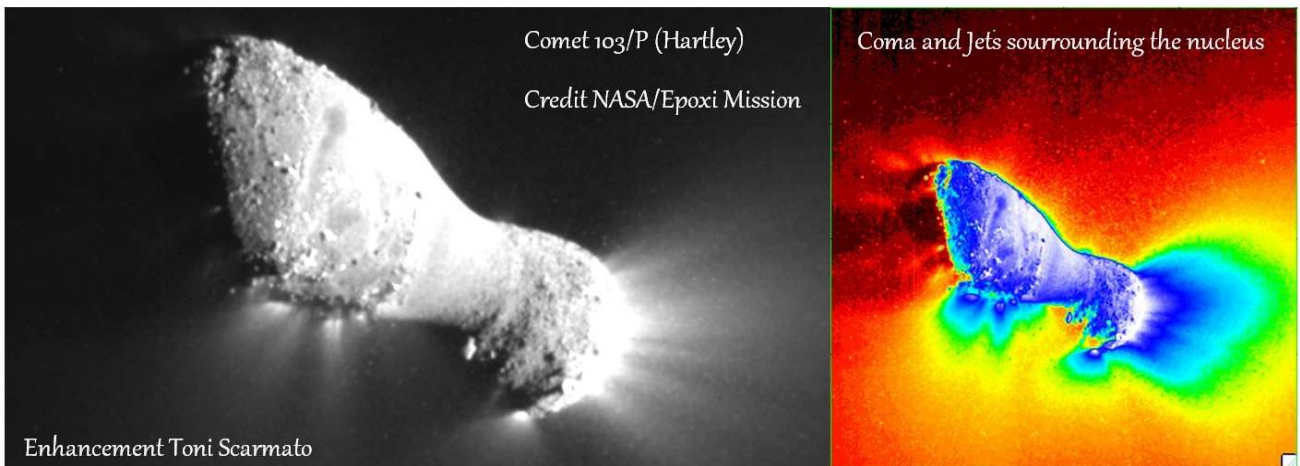

**Fig. 26 – Comet 103/P (Hartley) captured by Deep Impact Nasa/Epoxy Mission.**

To understand better the method to separate the contribution of the nucleus from the coma brightness in images at low resolution (mean that the size of the nucleus is much smaller than the spatial resolution of one pixel), look at following images.

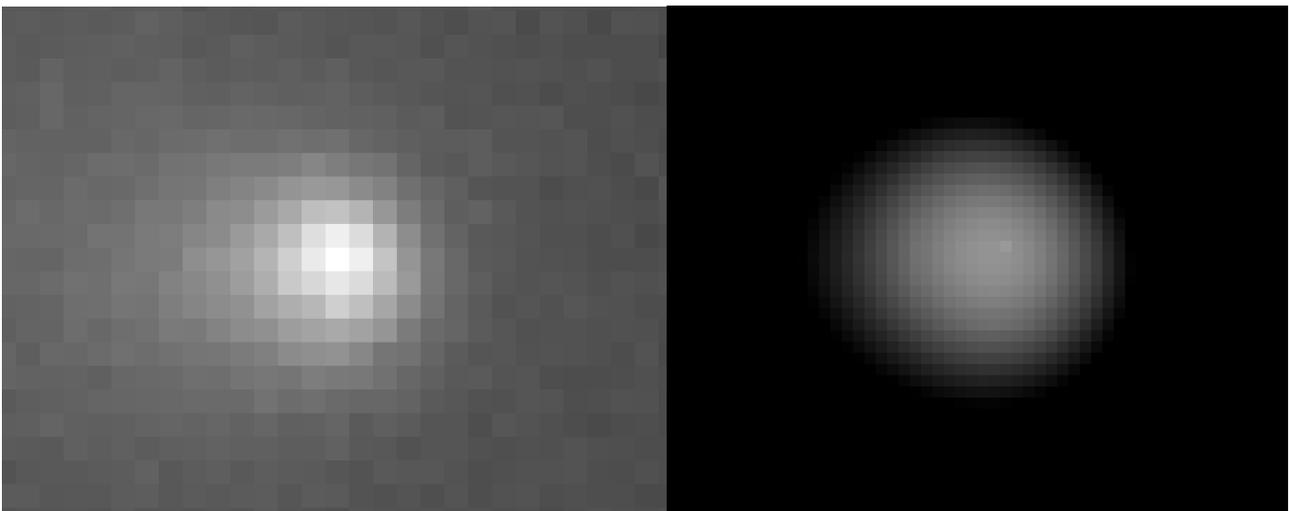

**Fig. 27 – This procedure is called "pixelization" ([Measuring the Size of ISON's Nucleus: Pixelization](); Jian-Yang Li; *Planetary Science Institute*). The image is relative to C/2011 L4 (PanStarrs) comet taken on 2013 July 29th. Left, crop of original image, 1,27 arcsec/pixel; right, crop of 4x4 resampled image, 0,08 arcsec/pixel.**

The procedure consist in the following steps; resample the image using **Bicubic Interpolation**, apply the **Convolution** and **PSF** (Point Spread Function). As described in (**Toni Scarmato 2014**, http://arxiv.org/abs/1405.3112)  the image at right in figure 27 was obtained in the following way. The 4x4 Bicubic Interpolation applied to the left image permit to obtain an image in which the value of an "x,y" point of one pixel is computed using the value of the 16 pixels around.   After this step need to apply the Convolution and PSF to obtain a new image corrected for any problems that affect it.  In particular, PSF is an algorithm that corrects problems due to the photons that overflows the pixels and create an image that is not real. The overflow of the photons can be due to the seeing, bad tracking and optical problems of the telescope.  Next step is construct a model that fits well the profile around the central pixel containing the nucleus and extrapolate, using Dirac Function, the contribution of the nucleus to the brightness.  This procedure work well until to five sub-pixels from the center and allows to define the power law that describe the profile of the coma around the nucleus.

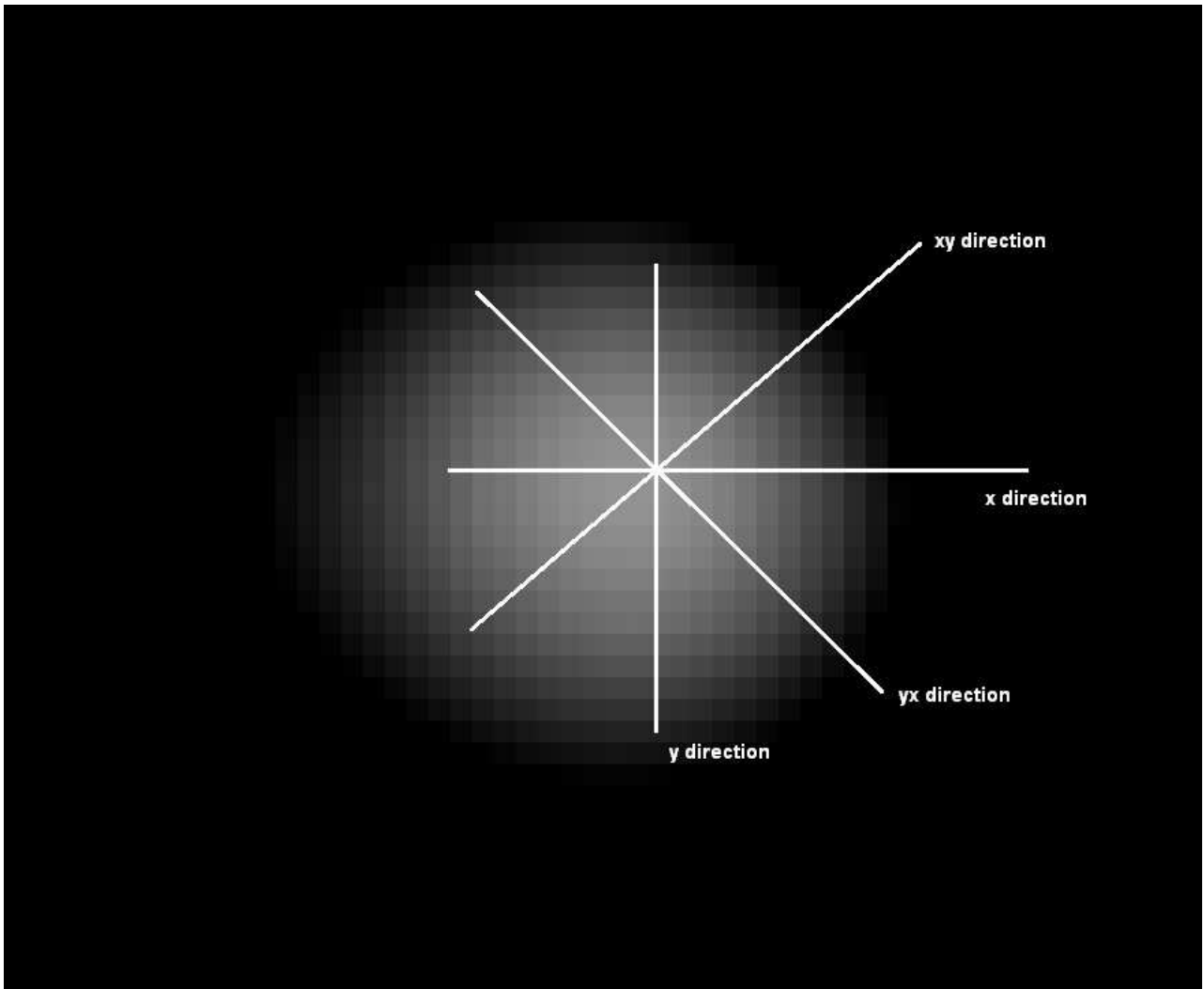

**Fig. 28 – Direction in the inner coma of the comet C/2011 L4 after the Pixelization used to plot the profiles around the nucleus and measure the contribution of the nucleus and coma to the total brightness.**

Image above show the four directions considered to define the trend of the coma. The central brighter pixel contain the contribution Coma+Nucleus. At the comet distance one pixel in our resampled image have a spatial resolution of 159 km. So if the point x0,y0 is the center of the brighter pixel and the position of the nucleus, we have from the x and y direction border 79,5 km, and from xy and yx direction border 112,4 km. In first approximation I can assume that the contribution of the nucleus is the same in the central pixel and in the surrounding pixels because the used procedure have reconstruct the x0,y0 nucleus position of the central pixel using the 16 pixels around after assigning zero value to the brightest central pixel. In this way, Dirac Function can extracts the residual value of the nucleus in agreement with Eqs. 7 and 8 in 3.1 above. Residual value in ADU is the brightness of the nucleus that, using differential photometry, it is converted in apparent magnitude of the nucleus and after in absolute magnitude, in agreement with Eq. 11 in 3.1 section. Now let us analyze the observations. First observation of the comet was on 2013, July 29$^{th}$. Results are summarized in the following tables and figures.

**Tab. 7** – **Results on 2013 July 29 observation along X axis**.

IMAGE RESAMPLED with 1/rho^a model - X **direction - 1pixel=159 km**

| X dir | ADU | model | Coma | X | Nucleus | X | delta |
|---|---|---|---|---|---|---|---|
| -5 | 27610 | 26980 | 26258 | -5 | 0 | -5 | 0.0000 |
| -4 | 28060 | 27602 | 26980 | -4 | 0 | -4 | 0.0000 |
| -3 | 28434 | 28117 | 27602 | -3 | 0 | -3 | 0.0000 |
| -2 | 28711 | 28516 | 28117 | -2 | 0 | -2 | 0.0000 |
| -1 | 28873 | 28788 | 28516 | -1 | 0 | -1 | 0.0000 |
| 0 | 28903 | 28903 | 28788 | 0 | 73 | 0 | 16311 |
| 1 | 28788 | 28788 | 28516 | 1 | 0 | 1 | 0.0000 |
| 2 | 28517 | 28516 | 28117 | 2 | 0 | 2 | 0.0000 |
| 3 | 28083 | 28117 | 27602 | 3 | 0 | 3 | 0.0000 |
| 4 | 27483 | 27602 | 26980 | 4 | 0 | 4 | 0.0000 |
| 5 | 27332 | 26980 | 26258 | 5 | 0 | 5 | 0.0000 |

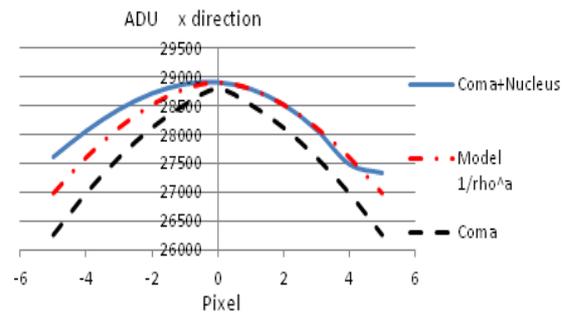

**Fig. 29 – Plot of the 2013 July 29th observation along X direction. Model fits well the measured curve using a power law with a=-1,75**

Table 7 show the ADU values along X direction starting from the central pixel. Figure 29 is the plot of the three curves. Coma+Nucleus is the plot of the profile as measured on the image, model is the plot based on the power law with **a=-1.75** that fits well the measured profile up to at least three pixels. One pixel is 159 km at the comet distance, so we have a good agreement of the two curve up to 308 km from the x0,y0 position. Dirac Function extract **73 ADU** as residuals in agreement with Eq.7 and Eq.8 in 3.1 section. This value correspond to **Ma=19,659+/-0,019** (magnitude apparent of the nucleus) that, considering the phase angle of 21,3° and a phase coefficient of 0,04 mag/deg, provides **H=18.807+/-0,019** (absolute magnitude of the nucleus) in agreement with Eq.11. Using Eq.10 in section 3.1 we have a comet radius equivalent to **Rn=495+/-87 m**.

**Tab. 8 - Results on 2013 July 29 observation along Y axis**.

IMAGE RESAMPLED with 1/rho^a model - Y **direction - 1pixel=159 km**

| Y dir | ADU | model | Coma | x | Nucleus | x | delta |
|---|---|---|---|---|---|---|---|
| -5 | 27174 | 26793 | 26695 | -5 | 0 | -5 | 0.0000 |
| -4 | 27758 | 27528 | 26793 | -4 | 0 | -4 | 0.0000 |
| -3 | 28244 | 28112 | 27528 | -3 | 0 | -3 | 0.0000 |
| -2 | 28610 | 28540 | 28112 | -2 | 0 | -2 | 0.0000 |
| -1 | 28834 | 28807 | 28540 | -1 | 0 | -1 | 0.0000 |
| 0 | 28903 | 28903 | 28834 | 0 | 83 | 0 | 16311 |
| 1 | 28807 | 28807 | 28540 | 1 | 0 | 1 | 0.0000 |
| 2 | 28540 | 28540 | 28112 | 2 | 0 | 2 | 0.0000 |
| 3 | 28109 | 28112 | 27528 | 3 | 0 | 3 | 0.0000 |
| 4 | 27800 | 27528 | 26793 | 4 | 0 | 4 | 0.0000 |
| 5 | 27204 | 26793 | 26695 | 5 | 0 | 5 | 0.0000 |

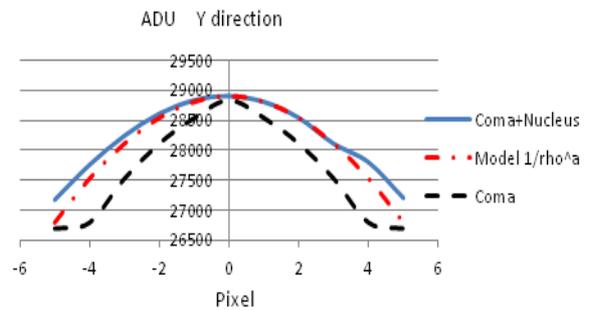

**Fig. 30 – Plot of the 2013 July 29th observation along Y direction. Model fits well the measured curve using a power law with a=-1,92**

Along Y axis the coma profile is not much different respect X axis, residual are **83 ADU** corresponding to **Ma=19,519+/-0.018** and **H=18,667+/-0,018** with a comet radius **Rn=528+/-93 m**. Power law is depending by the factor **a=-1,92**. Results along respectively XY and YX axis give the following data; residual 143 and 167 ADU, **Ma=18,922+/-0.021** and **H=18,070+/-0,020** for XY direction, **Ma=18,757+/-0,020** and **H=18,070+/-0,020** for YX, and the radius of the comet equal to **Rn=695+/-123 m** and **Rn=750+/-132 m..** Power exponent are **a=-2,14** and **a=-1,84** respectively for XY and YX direction.

**Tab. 9 -** Results on 2013 July 29 observation along XY axis.

IMAGE RESAMPLED with 1/rho model - XY direction - 1pixel=159 km

| XY | ADU | model | Coma | x | Nucleus | x | delta |
|---|---|---|---|---|---|---|---|
| -5 | 25839 | 25270 | 23439 | -5 | 0 | -5 | 0.0000 |
| -4 | 26668 | 26649 | 25270 | -4 | 0 | -4 | 0.0000 |
| -3 | 27588 | 27685 | 26649 | -3 | 0 | -3 | 0.0000 |
| -2 | 28286 | 28392 | 27685 | -2 | 0 | -2 | 0.0000 |
| -1 | 28733 | 28787 | 28392 | -1 | 0 | -1 | 0.0000 |
| 0 | 28903 | 28903 | 28733 | 0 | 143 | 0 | 16311 |
| 1 | 28787 | 28787 | 28392 | 1 | 0 | 1 | 0.0000 |
| 2 | 28391 | 28392 | 27685 | 2 | 0 | 2 | 0.0000 |
| 3 | 27747 | 27685 | 26649 | 3 | 0 | 3 | 0.0000 |
| 4 | 26903 | 26649 | 25270 | 4 | 0 | 4 | 0.0000 |
| 5 | 26089 | 25270 | 23439 | 5 | 0 | 5 | 0.0000 |

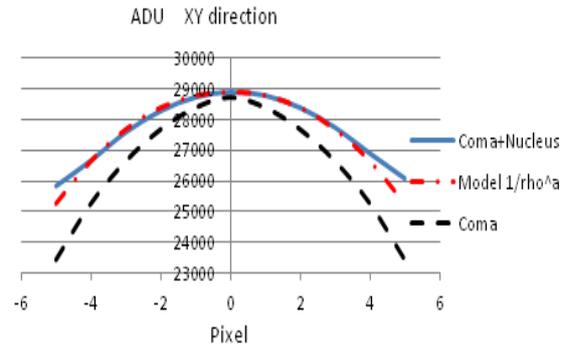

Fig. 31 – Plot of the 2013 July 29th observation along XY direction. Model fits well the measured curve using a power law with a=-2,14

**Tab. 10 -** Results on 2013 July 29 observation along YX axis.

IMAGE RESAMPLED with 1/rho model - YX direction - 1pixel=159 km

| YX | ADU | Model | Coma | pixel | Nucleus | pixel | delta |
|---|---|---|---|---|---|---|---|
| -5 | 26088 | 24536 | 22795 | -5 | 0 | -5 | 0.0000 |
| -4 | 26995 | 26006 | 24536 | -4 | 0 | -4 | 0.0000 |
| -3 | 27779 | 27197 | 26006 | -3 | 0 | -3 | 0.0000 |
| -2 | 28401 | 28094 | 27197 | -2 | 0 | -2 | 0.0000 |
| -1 | 28796 | 28677 | 28094 | -1 | 0 | -1 | 0.0000 |
| 0 | 28903 | 28903 | 28677 | 0 | 167 | 0 | 16311 |
| 1 | 28677 | 28677 | 28094 | 1 | 0 | 1 | 0.0000 |
| 2 | 28090 | 28094 | 27197 | 2 | 0 | 2 | 0.0000 |
| 3 | 27144 | 27197 | 26006 | 3 | 0 | 3 | 0.0000 |
| 4 | 25872 | 26006 | 24536 | 4 | 0 | 4 | 0.0000 |
| 5 | 25084 | 24536 | 22795 | 5 | 0 | 5 | 0.0000 |

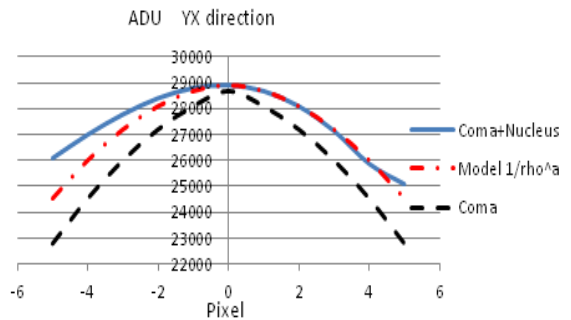

Fig. 32 – Plot of the 2013 July 29th observation along YX direction. Model fits well the measured curve using a power law with a=-1,84

All the data are summarized in the table 11 here below with the average values.

**COMET C/2011 L4 (PANSTARRS) - 2013 July 29th - Toni Scarmato's Observatory**

Table 11    Band R    albedo p=0,04    1 A.U.=149,577,000 km

| axis | ADU | magapp | H (absolute) | Rcomet (m) | err +/- | a (rho^a) | Phase ° | D (AU) | r (AU) |
|---|---|---|---|---|---|---|---|---|---|
| x | 73 | 19.659 | 18.807 | 495 | 87 | -1.75 | 21.3 | 2.769 | 2.709 |
| y | 83 | 19.519 | 18.667 | 528 | 93 | -1.92 | 21.3 | 2.769 | 2.709 |
| xy | 143 | 18.922 | 18.070 | 695 | 123 | -2.14 | 21.3 | 2.769 | 2.709 |
| yx | 167 | 18.757 | 17.905 | 750 | 132 | -1.84 | 21.3 | 2.769 | 2.709 |
| Average | 116 | 19.148 | 18.296 | 617 | 109 | -1.91 | 21.3 | 2.769 | 2.709 |

density  0.6  g*cm^-3  Ag(gravitational accelaration)  0.010  m*s^-2
mass  5.90E+11  kg  Ve(escape velocity)  0.252  m*s^-1

On **2013 August 1th**, another good night to observe the comet, it has allowed me to get another time-series of data.

**Tab. 12 -** Results on 2013 August 1th observation along X axis.

IMAGE RESAMPLED with 1/rho model - X direction - 1pixel=163 km

| X dir | ADU | model | Coma | X | Nucleus | X | delta |
|---|---|---|---|---|---|---|---|
| -5 | 15645 | 15545 | 14878 | -5 | 0 | -5 | 0.0000 |
| -4 | 15908 | 16002 | 15545 | -4 | 0 | -4 | 0.0000 |
| -3 | 16135 | 16285 | 16002 | -3 | 0 | -3 | 0.0000 |
| -2 | 16316 | 16436 | 16285 | -2 | 0 | -2 | 0.0000 |
| -1 | 16441 | 16494 | 16409 | -1 | 0 | -1 | 0.0000 |
| 0 | 16503 | 16503 | 16494 | 0 | 36 | 0 | 9313 |
| 1 | 16494 | 16494 | 16409 | 1 | 0 | 1 | 0.0000 |
| 2 | 16409 | 16436 | 16285 | 2 | 0 | 2 | 0.0000 |
| 3 | 16246 | 16285 | 16002 | 3 | 0 | 3 | 0.0000 |
| 4 | 16005 | 16002 | 15545 | 4 | 0 | 4 | 0.0000 |
| 5 | 15689 | 15545 | 14878 | 5 | 0 | 5 | 0.0000 |

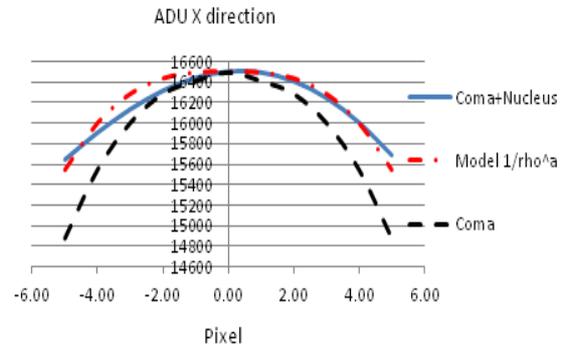

Fig. 33 – Plot of the 2013 August 1th observation along X direction. Model fits well the measured curve using a power law with a=-2,90

**Tab. 13 -** Results on 2013 August 1th observation along Y axis.

IMAGE RESAMPLED with 1/rho model - Y direction - 1pixel=163 km

| Y dir | ADU | model | Coma | x | Nucleus | x | delta |
|---|---|---|---|---|---|---|---|
| -5 | 15500 | 15266 | 14801 | -5 | 0 | -5 | 0.0000 |
| -4 | 15862 | 15666 | 15266 | -4 | 0 | -4 | 0.0000 |
| -3 | 16153 | 15997 | 15666 | -3 | 0 | -3 | 0.0000 |
| -2 | 16361 | 16254 | 15997 | -2 | 0 | -2 | 0.0000 |
| -1 | 16479 | 16429 | 16254 | -1 | 0 | -1 | 0.0000 |
| 0 | 16503 | 16503 | 16429 | 0 | 49 | 0 | 9313 |
| 1 | 16429 | 16429 | 16254 | 1 | 0 | 1 | 0.0000 |
| 2 | 16257 | 16254 | 15997 | 2 | 0 | 2 | 0.0000 |
| 3 | 15989 | 15997 | 15666 | 3 | 0 | 3 | 0.0000 |
| 4 | 15630 | 15666 | 15266 | 4 | 0 | 4 | 0.0000 |
| 5 | 15372 | 15266 | 14801 | 5 | 0 | 5 | 0.0000 |

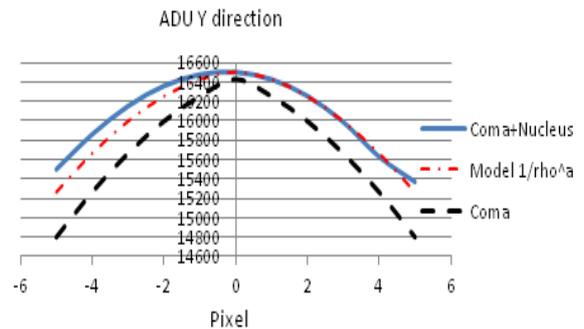

Fig. 34 – Plot of the 2013 August 1th observation along Y direction. Model fits well the measured curve using a power law with a=-1,75

It is interesting to note that, in the "negative" direction, the model don't fits well the measured profile. This "effect" look like relevant after the second pixel so at about 330 km from the nucleus. It is also possible to note that we have almost the same trend on **2013 July 29th**. At the moment we can say that –X direction is that of the tail so we can to suppose that the coma is really brighter in that direction. Further analysis will be done in the next section. Measured magnitude along X direction, with 36 ADU residual, phase angle of 20,9° and phase coefficient 0,04 ma/deg, is **Ma=20,596+/-0.021**, **H=19,760+/-0,021** and **Rn=319+/-56 m** with **a=-2,90.** Along Y axis **Ma=20,246/-0,023**, **H=19,410/-0,022**, **Rn=375/-66 m** and exponent **a=-1,75**, residual ADU are 49. Along other two axis XY and YX we have **Ma=19,648+/-0,23** and **H=18,812+/-0,23, Rn=494+/-87 m** and **a=-1,65, Ma=19,674+/-0,21, H=18,838+/-0,21** and **Rn=488+/-86 m, a=-2,05** with an average values **Ra=419+/-74 m** and **a=-2,09.** All the results are summarized here in the tables and figures below.

## Tab. 14 - Results on 2013 August 1th observation along XY axis

IMAGE RESAMPLED with 1/rho model - XY **direction - 1pixel=163 km**

| XY | ADU | model | Coma | x | Nucleus | x | delta |
|---|---|---|---|---|---|---|---|
| -5 | 14676 | 14582 | 13907 | -5 | 0 | -5 | 0.0000 |
| -4 | 15337 | 15173 | 14582 | -4 | 0 | -4 | 0.0000 |
| -3 | 15874 | 15676 | 15173 | -3 | 0 | -3 | 0.0000 |
| -2 | 16257 | 16079 | 15676 | -2 | 0 | -2 | 0.0000 |
| -1 | 16468 | 16368 | 16079 | -1 | 0 | -1 | 0.0000 |
| 0 | 16503 | 16503 | 16368 | 0 | 85 | 0 | 9313 |
| 1 | 16368 | 16368 | 16079 | 1 | 0 | 1 | 0.0000 |
| 2 | 16082 | 16079 | 15676 | 2 | 0 | 2 | 0.0000 |
| 3 | 15667 | 15676 | 15173 | 3 | 0 | 3 | 0.0000 |
| 4 | 15154 | 15173 | 14582 | 4 | 0 | 4 | 0.0000 |
| 5 | 14665 | 14582 | 13907 | 5 | 0 | 5 | 0.0000 |

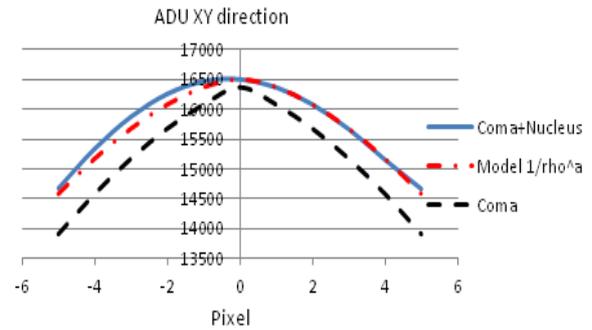

Fig. 34 – Plot of the 2013 August 1th observation along XY direction. Model fits well the measured curve using a power law with a=-1,65

## Tab. 15 - Results on 2013 August 1th observation along YX axis

IMAGE RESAMPLED with 1/rho model - YX **direction - 1pixel=163 km**

| YX | ADU | Model | Coma | x | Nucleus | x | delta |
|---|---|---|---|---|---|---|---|
| -5 | 14879 | 14254 | 13235 | -5 | 0 | -5 | 0.0000 |
| -4 | 15393 | 15080 | 14254 | -4 | 0 | -4 | 0.0000 |
| -3 | 15839 | 15714 | 15080 | -3 | 0 | -3 | 0.0000 |
| -2 | 16190 | 16159 | 15714 | -2 | 0 | -2 | 0.0000 |
| -1 | 16420 | 16420 | 16159 | -1 | 0 | -1 | 0.0000 |
| 0 | 16503 | 16503 | 16420 | 0 | 83 | 0 | 9313 |
| 1 | 16420 | 16420 | 16159 | 1 | 0 | 1 | 0.0000 |
| 2 | 16157 | 16159 | 15714 | 2 | 0 | 2 | 0.0000 |
| 3 | 15714 | 15714 | 15080 | 3 | 0 | 3 | 0.0000 |
| 4 | 15319 | 15080 | 14254 | 4 | 0 | 4 | 0.0000 |
| 5 | 14551 | 14254 | 13235 | 5 | 0 | 5 | 0.0000 |

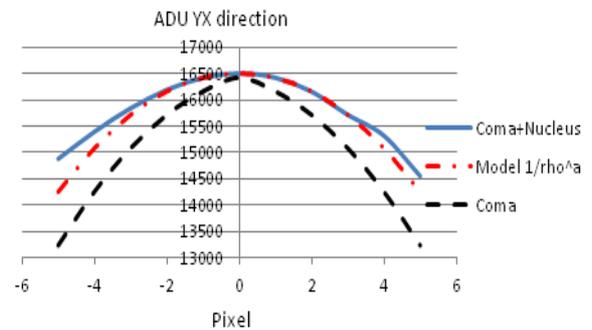

Fig. 35 – Plot of the 2013 August 1th observation along XY direction. Model fits well the measured curve using a power law with a=-2,05

### COMET C/2011 L4 (PANSTARRS) - 2013 August 1th - Toni Scarmato's Observatory

Table 16      Band R    albedo p=0,04                    1 A.U.=149,577,000 km

| axis | ADU | magapp | H (absolute) | Rcomet (m) | err +/- | a (rho^a) | Phase ° | D (AU) | r (AU) |
|---|---|---|---|---|---|---|---|---|---|
| x | 36 | 20.596 | 19.760 | 319 | 56 | -2.90 | 20.9 | 2.825 | 2.750 |
| y | 49 | 20.246 | 19.410 | 375 | 66 | -1.75 | 20.9 | 2.825 | 2.750 |
| xy | 85 | 19.648 | 18.812 | 494 | 87 | -1.65 | 20.9 | 2.825 | 2.750 |
| yx | 83 | 19.674 | 18.838 | 488 | 86 | -2.05 | 20.9 | 2.825 | 2.750 |
| Average | 63 | 19.971 | 19.135 | 419 | 74 | -2.09 | 20.9 | 2.825 | 2.750 |

| | | | | | | |
|---|---|---|---|---|---|---|
| density | 0.6 | g*cm^-3 | Ag(gravitational accelaration) | 0.008 | m*s^-2 | |
| mass | 1.85E+11 | kg | Ve(escape velocity ) | 0.171 | m*s^-1 | |

Finally, third night of observation, on **2013 August 3th**, the comet was at about 3 A.U. from the Earth and Sun, yet active with an Af(rho) value of about 4300 cm, equivalent to 3,72x10^8 kg/day of dust lost (see table 6). The previous measures show that the nucleus of the comet is small but really very active in the production of dust. Here in the table below are summarized the data of the latest measure that confirm C/2011 L4 be a small comet. Phase angle equal to 20,7° is not much different since the previous observations. Along X direction we have **Ma=20,840+/-0,025, H=20,012+/-0,024, Rn=284+/-50 m** with exponent **a=-2,46.** Along Y axis **Ma=20,420+/-0,018, H=19,592+/-0,019, Rn=345+/-61 m** and **a=-1,75.** Regarding other two axis we have the same values, **Ma=19,858+/-0,021, H=19,030+/-0,021, Rn=447+/-79 m,** only the exponents of the power law are different, respectively **a=-1,65** for XY direction and **a=-1,85** for YX. The average values are, **Ma=20,170+/-0,020, H=19,342+/-0,020** and corresponding radius **Rn=381+/-67m** and exponent **a=-1,93.** Summary in the following tables and plots**.**

**Tab. 17 -** Results on 2013 August 1th observation along X axis
IMAGE RESAMPLED with 1/rho model - X **direction - 1pixel=165 km**

| X dir | ADU | model | Coma | X | Nucleus | X | delta |
|---|---|---|---|---|---|---|---|
| -5.00 | 12811 | 12679 | 12234 | -5 | 0 | -5 | 0.0000 |
| -4.00 | 13021 | 13011 | 12679 | -4 | 0 | -4 | 0.0000 |
| -3.00 | 13197 | 13241 | 13011 | -3 | 0 | -3 | 0.0000 |
| -2.00 | 13334 | 13382 | 13241 | -2 | 0 | -2 | 0.0000 |
| -1.00 | 13425 | 13450 | 13380 | -1 | 0 | -1 | 0.0000 |
| 0.00 | 13465 | 13465 | 13450 | 0 | 28 | 0 | 7599 |
| 1.00 | 13450 | 13450 | 13380 | 1 | 0 | 1 | 0.0000 |
| 2.00 | 13377 | 13382 | 13241 | 2 | 0 | 2 | 0.0000 |
| 3.00 | 13242 | 13241 | 13011 | 3 | 0 | 3 | 0.0000 |
| 4.00 | 13045 | 13011 | 12679 | 4 | 0 | 4 | 0.0000 |
| 5.00 | 12787 | 12679 | 12234 | 5 | 0 | 5 | 0.0000 |

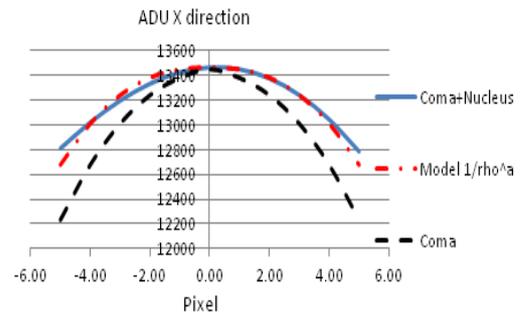

Fig. 36 – Plot of the 2013 August 3th observation along X direction. Model fits well the measured curve using a power law with a=-2,46

**Tab. 18 -** Results on 2013 August 1th observation along Y axis
IMAGE RESAMPLED with 1/rho model - Y **direction - 1pixel=165 km**

| Y dir | ADU | model | Coma | x | Nucleus | x | delta |
|---|---|---|---|---|---|---|---|
| -5 | 12686 | 12261 | 11809 | -5 | 0 | -5 | 0.0000 |
| -4 | 12975 | 12650 | 12261 | -4 | 0 | -4 | 0.0000 |
| -3 | 13205 | 12973 | 12650 | -3 | 0 | -3 | 0.0000 |
| -2 | 13368 | 13223 | 12973 | -2 | 0 | -2 | 0.0000 |
| -1 | 13456 | 13393 | 13223 | -1 | 0 | -1 | 0.0000 |
| 0 | 13465 | 13465 | 13393 | 0 | 41 | 0 | 7599 |
| 1 | 13393 | 13393 | 13223 | 1 | 0 | 1 | 0.0000 |
| 2 | 13240 | 13223 | 12973 | 2 | 0 | 2 | 0.0000 |
| 3 | 13013 | 12973 | 12650 | 3 | 0 | 3 | 0.0000 |
| 4 | 12893 | 12650 | 12261 | 4 | 0 | 4 | 0.0000 |
| 5 | 12532 | 12261 | 11809 | 5 | 0 | 5 | 0.0000 |

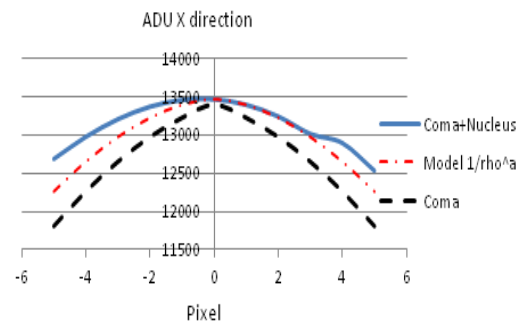

Fig. 37 – Plot of the 2013 August 3th observation along Y direction. Model fits well the measured curve using a power law with a=-1,75

**Tab. 19 -** Results on 2013 August 1th observation along XY axis

IMAGE RESAMPLED with 1/rho model - XY **direction - 1pixel=165 km**

| XY | ADU | model | Coma | x | Nucleus | x | delta |
|---|---|---|---|---|---|---|---|
| -5 | 12039 | 11871 | 11311 | -5 | 0 | -5 | 0.0000 |
| -4 | 12558 | 12362 | 11871 | -4 | 0 | -4 | 0.0000 |
| -3 | 12978 | 12779 | 12362 | -3 | 0 | -3 | 0.0000 |
| -2 | 13276 | 13114 | 12779 | -2 | 0 | -2 | 0.0000 |
| -1 | 13441 | 13353 | 13114 | -1 | 0 | -1 | 0.0000 |
| 0 | 13465 | 13465 | 13353 | 0 | 68 | 0 | 7599 |
| 1 | 13353 | 13353 | 13114 | 1 | 0 | 1 | 0.0000 |
| 2 | 13117 | 13114 | 12779 | 2 | 0 | 2 | 0.0000 |
| 3 | 12777 | 12779 | 12362 | 3 | 0 | 3 | 0.0000 |
| 4 | 12480 | 12362 | 11871 | 4 | 0 | 4 | 0.0000 |
| 5 | 11985 | 11871 | 11311 | 5 | 0 | 5 | 0.0000 |

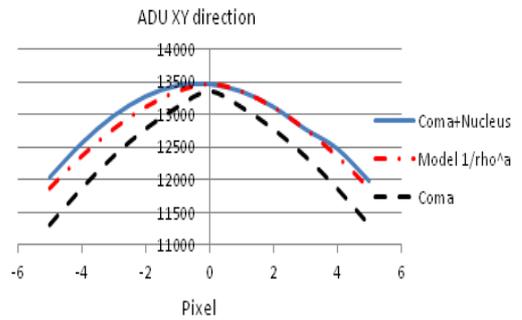

**Fig. 38** – Plot of the 2013 August 3th observation along XY direction. Model fits well the measured curve using a power law with a=-1,65

**Tab. 20 -** Results on 2013 August 1th observation along YX axis

IMAGE RESAMPLED with 1/rho model - YX **direction - 1pixel=165 km**

| YX | ADU | Model | Coma | x | Nucleus | x | delta |
|---|---|---|---|---|---|---|---|
| -5 | 12166 | 11737 | 11044 | -5 | 0 | -5 | 0.0000 |
| -4 | 12598 | 12321 | 11737 | -4 | 0 | -4 | 0.0000 |
| -3 | 12965 | 12793 | 12321 | -3 | 0 | -3 | 0.0000 |
| -2 | 13244 | 13148 | 12793 | -2 | 0 | -2 | 0.0000 |
| -1 | 13417 | 13377 | 13148 | -1 | 0 | -1 | 0.0000 |
| 0 | 13465 | 13465 | 13377 | 0 | 68 | 0 | 7599 |
| 1 | 13377 | 13377 | 13148 | 1 | 0 | 1 | 0.0000 |
| 2 | 13148 | 13148 | 12793 | 2 | 0 | 2 | 0.0000 |
| 3 | 12781 | 12793 | 12321 | 3 | 0 | 3 | 0.0000 |
| 4 | 12469 | 12321 | 11737 | 4 | 0 | 4 | 0.0000 |
| 5 | 11855 | 11737 | 11044 | 5 | 0 | 5 | 0.0000 |

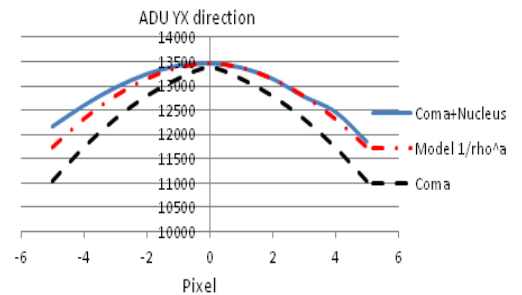

**Fig. 39** – Plot of the 2013 August 3th observation along YX direction. Model fits well the measured curve using a power law with a=-1,85

| COMET C/2011 L4 (PanStarrs) - 2013 August 3 - Toni Scarmato's Observatory | | | | | | | | | |
|---|---|---|---|---|---|---|---|---|---|
| Table 21 | | Band R | albedo p=0,04 | | | 1 A.U.=149,577,000 km | | | |
| axis | ADU | magapp | H (absolute) | Rcomet (m) | err +/- | a (rho^a) | Phase ° | D (AU) | r (AU) |
| x | 28 | 20.840 | 20.012 | 284 | 50 | -2.46 | 20.7 | 2.863 | 2.778 |
| y | 41 | 20.420 | 19.592 | 345 | 61 | -1.75 | 20.7 | 2.863 | 2.778 |
| xy | 68 | 19.858 | 19.030 | 447 | 79 | -1.65 | 20.7 | 2.863 | 2.778 |
| yx | 68 | 19.858 | 19.030 | 447 | 79 | -1.85 | 20.7 | 2.863 | 2.778 |
| Average | 51 | 20.170 | 19.342 | 381 | 67 | -1.93 | 20.7 | 2.863 | 2.778 |
| density | 0.6 | g*cm^-3 | Ag(gravitational accelaration) | | 0.008 | m*s^-2 | | | |
| mass | 1.38E+11 | kg | Ve(escape velocity) | | 0.156 | m*s^-1 | | | |

## 5) DUST TAIL AND INNER COMA EVOLUTION

### 5.1 Dust tail and anti-tail

The evolution of the light curve of a comet in general is accompanied from the evolution of the tails and the coma. In the case of the comet C / 2011 L4 I have observed only a tail clearly, that of dust. In the days after perihelion, the dust tail was fanned. This would suggest the formation of an anti-tail when the Earth would cross the comet's orbital plane. The geometric situation is represented in the following graphs.

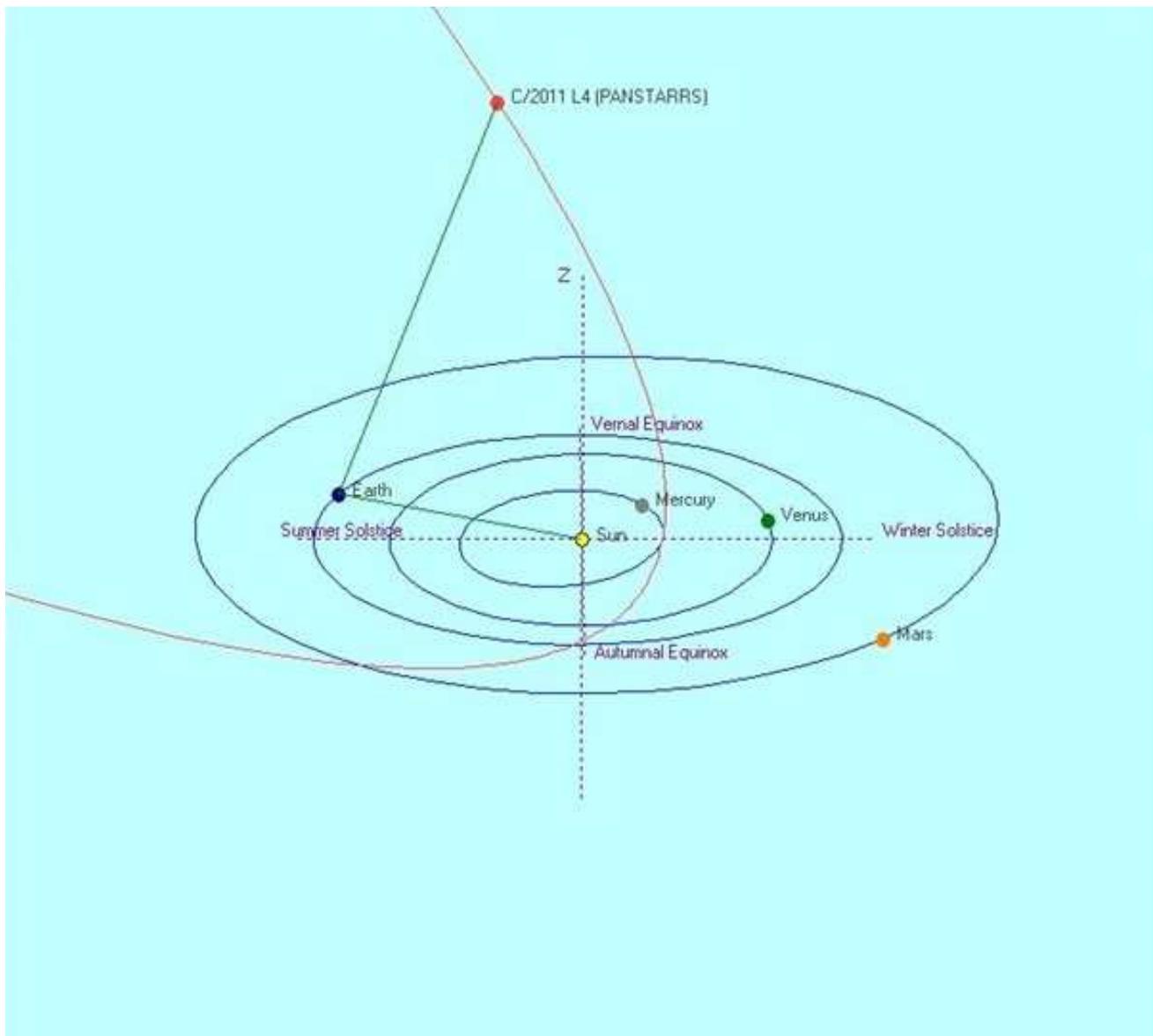

**Fig. 40 – Graph of the situation at the time of the Earth's cross of the orbital plane of the comet C/2011 L4.**

On **2013 May 26**[th] the Earth crossed the orbital plane of the comet. The observation of the comet has revealed a beautiful anti-tail. My observation on 2013 May 27[th] (see fig. 41) show the initial part of the tail pointed apparently toward the Sun. The dust released from the comet along the orbit, it is seen, from an observer terrestrial, like a tail that follows the comet and pointing towards the sun, because when the earth crosses the orbital plane, all the dust is observed perpendicularly along the line of sight. Then, an anti-tail is an "apparent" spike outgoing from a comet's coma that 3542a82f6a7d9887fb4cenAstronomy / CometsComet C/2011 L4 (PanStarrs)T. ScarmatoToni Scarmato's Observatory, Calabria, Italy2013report appear to point towards the Sun. Geometrically it is in opposite direction with respect to the other tails. This phenomenon is "apparent" and not real. The larger dust particles, that are affected mainly by the gravitational force of the Sun, follow the comet and can to form a disc along the comet's orbit. As Earth crosses the comet's orbital plane, the disc appears like the characteristic spike visible in figure 41.

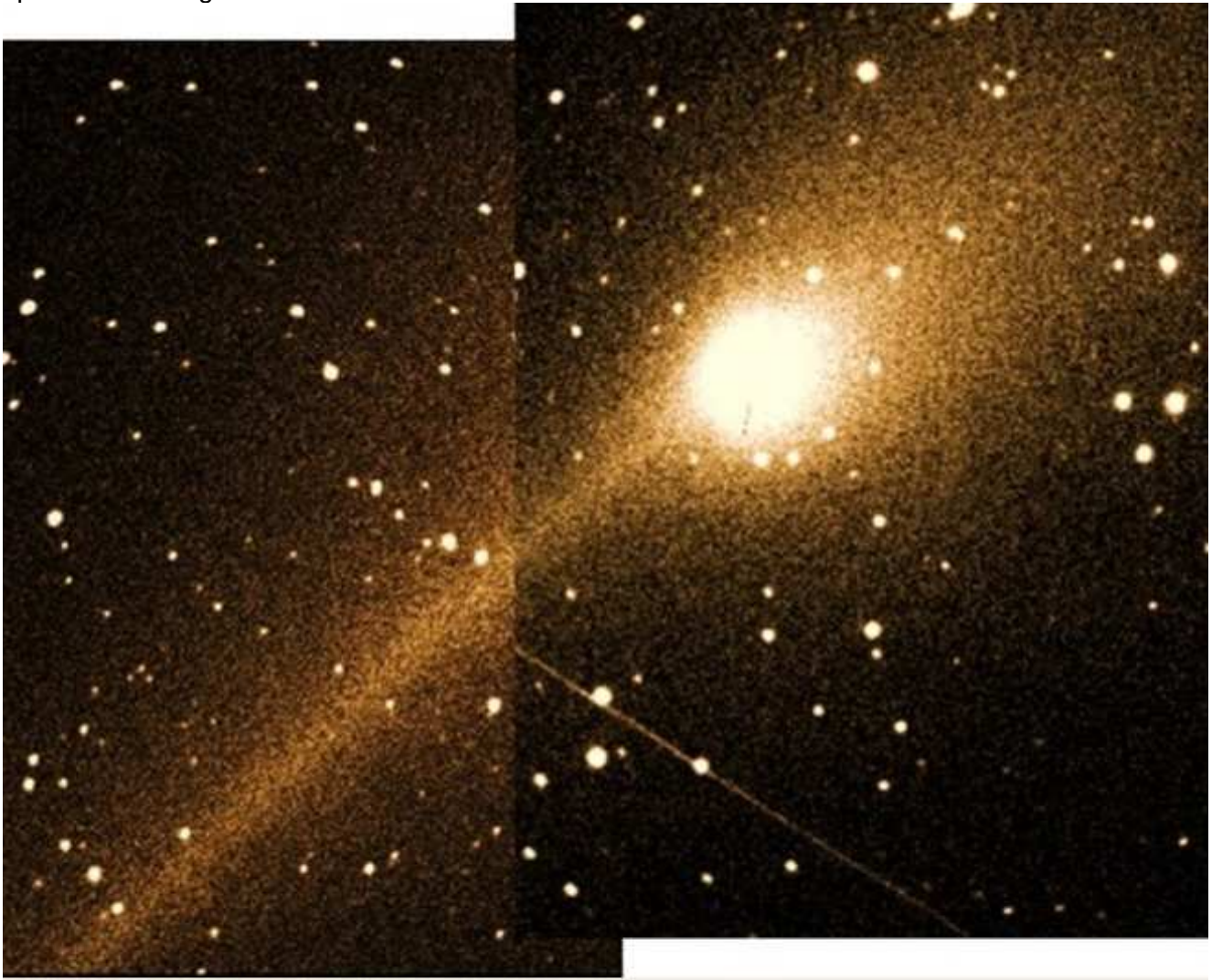

**Fig. 41 – Anti-tail in comet C/2011 L4 imaged in Rc band.**

I observe a very long anti-tail the night on **2013 June 2th**, using a large field image. The apparent tail toward the Sun was long about 4° that, at the comet distance, is equal to about 19,4 millions of kilometers (see Figure 42)

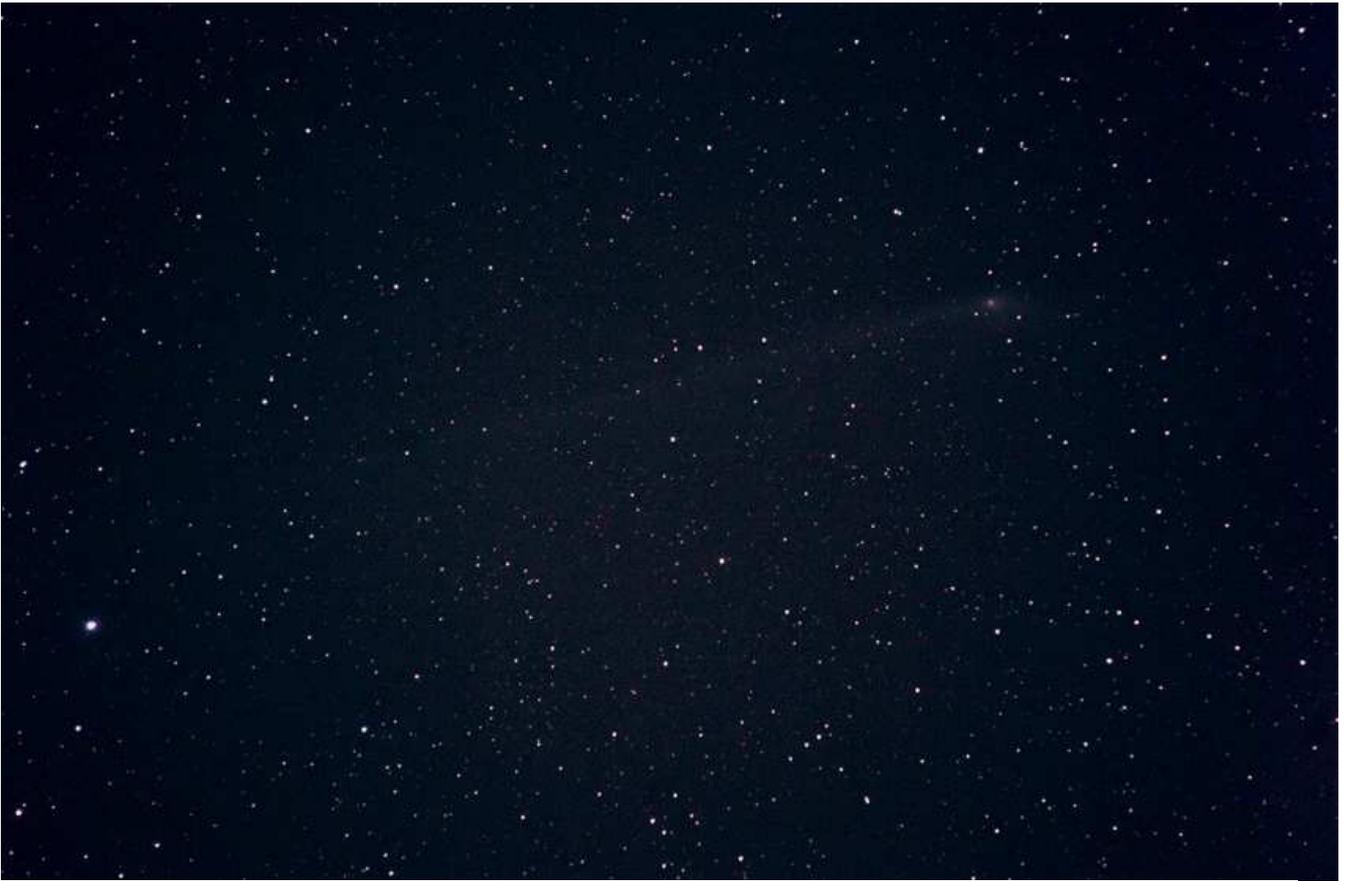

**Fig. 42 – The very long anti-tail of comet C/2011 L4 images on 2013 June 2th with Tele Apo 70 mm and Canon 10D.**

## 5.2 Inner coma structure and period of rotation

Another interesting analysis concerns the inner coma. The process of the images is based on two powerful filter in Astroart 3.0, **RWM (Radial Weighted Model)** and **MCM (Median Coma Model)**. 1/rho^a model and background subtraction is applied after having calibrated the original image with dark, flat and bias if necessary. The algorithms based on **Bonev, T. & Jockers, K. "Spatial distribution of the dust color in comet C/LINEAR (2000 WM1)"** (Proceedings of Asteroids, Comets, Meteors - ACM 2002. International Conference, 29 July - 2 August 2002, Berlin, Germany. Ed. Barbara Warmbein. ESA SP-500. Noordwijk, Netherlands: ESA Publications Division, ISBN 92-9092-810-7, 2002, p. 587 – 591). These algorithms extract the pixels value of the coma, subtracts the background value and multiplies for the cometcenter distance to create a new image with the computed values. After this, we used a crop of 40x40 pixels image centered on nucleus position, to apply the Bicubic Interpolation, Convolution and PSF. As said in Section 4, the sub-pixelization procedure permit also to identify the finer **GRADIENT OF BRIGHTNESS** in the coma and tail. Results obtained on the images of 7 nights of observation show interesting structures in the inner coma and clearly that the nucleus is a fast rotator in agreement with the result obtained with photometric analysis. On **2013 April 19$^{th}$** , 25 stacked images in Rc band, 120 second of exposure, show the main dust tail in PA~ 325° and the forming anti-tail in PA~125° (see Figure 43). As reported in table 6 the magnitude of the comet was 7,561+/- and Af(rho)=34249+/-513 cm for 17 arcsec of aperture. Then, the comet was in activity with a strong dust production. My elaboration of the inner coma show, in fact, an interesting structure linked with the dust production. Both RWM and MCM filters put in evidence the same beaviour with a "jet" like structure in PA~180° curved toward West (see figure 44).

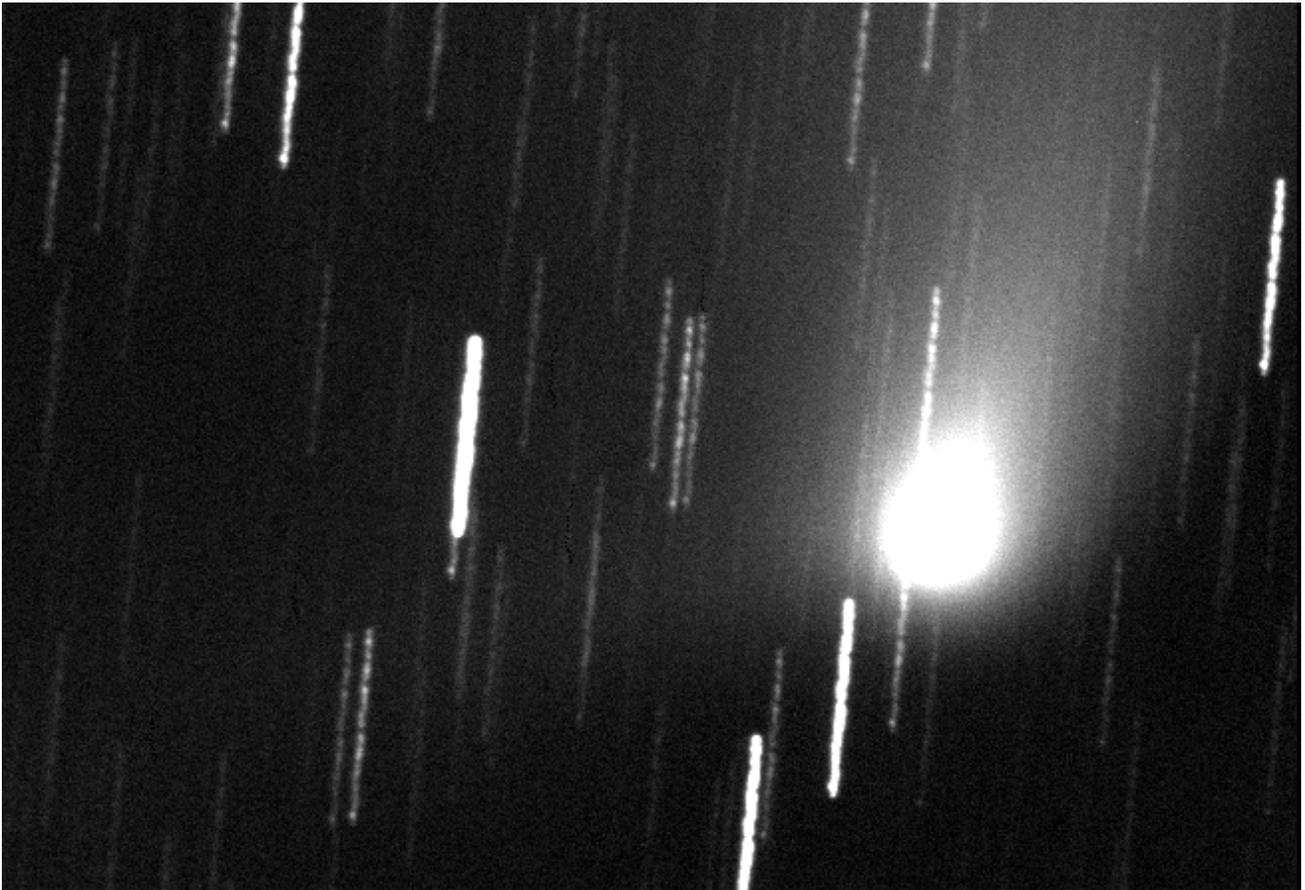

Fig. 43 – 2013 April 19th, 25x120 sec original CCD image, 25 cm Newton f/4.8 and CCD Atik 16 Ic with Rc filter. Start sequence 02:33:15 U.T.

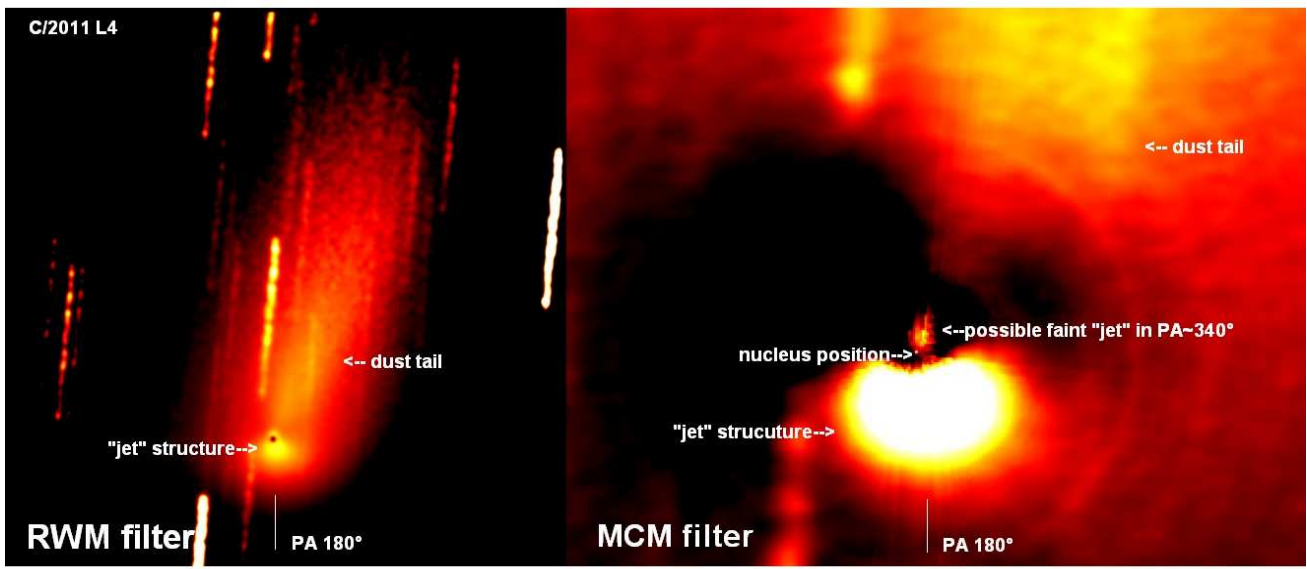

Fig. 44 – 2013 April 19th, left RWM filter, right MCM filter.

The presence of the two structures in figure 44 was confirmed by the observations in the days following. In particular the faint "jet" like structure is clearly rotating in two images on **2013 May 27th** taken on about 27 minutes apart. The evolution of the inner coma it can be seen in the next images.

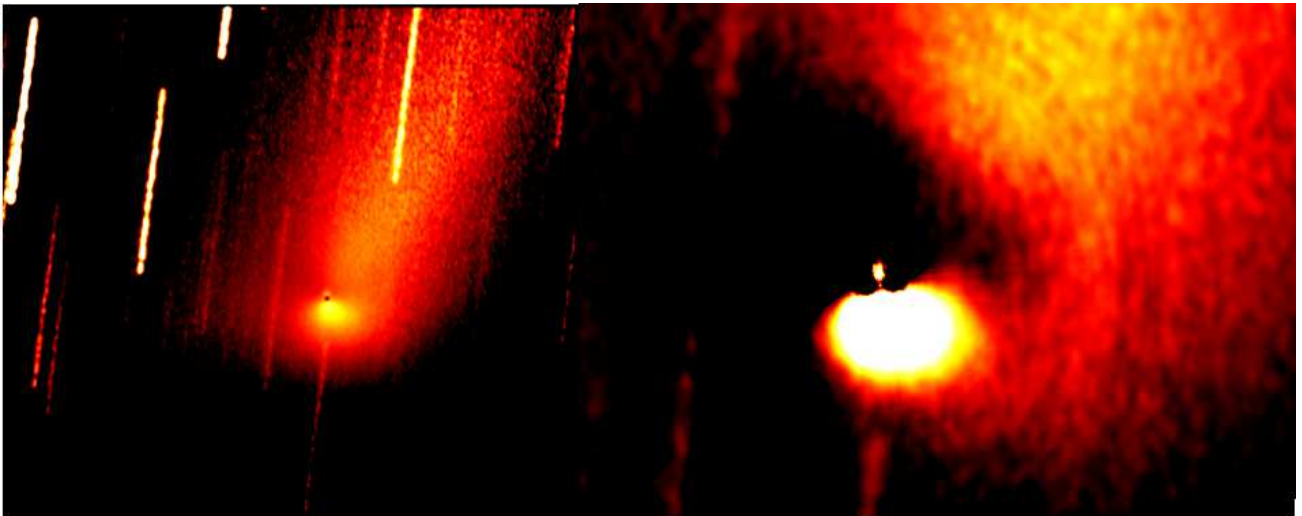

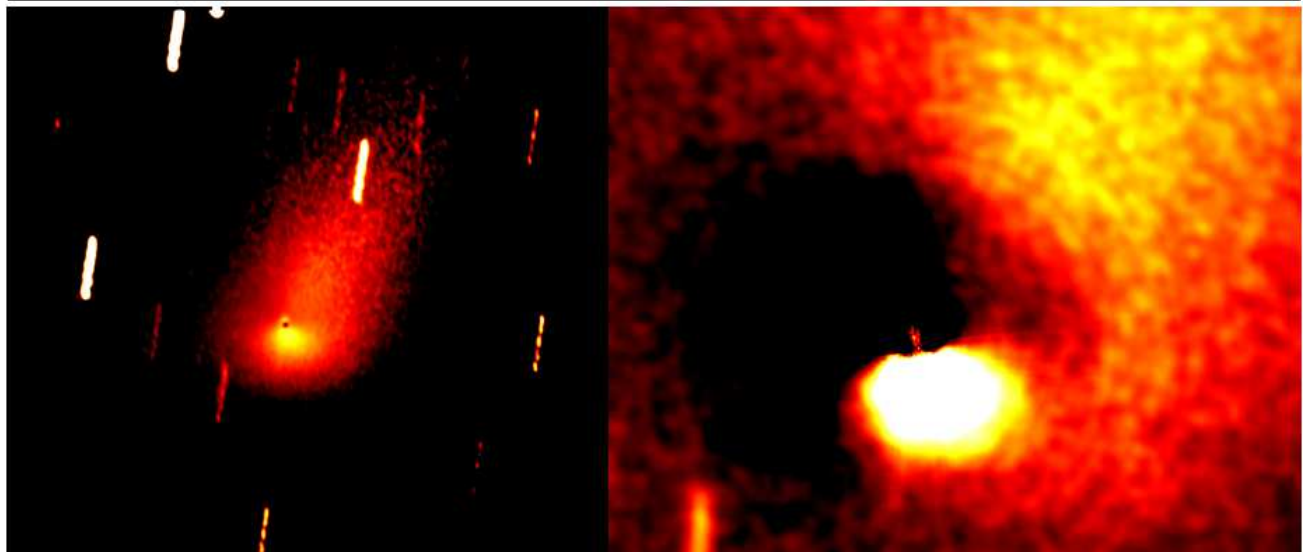

**Fig. 45 – 2013 April 28th,  left RWM filter, right MCM filter.**

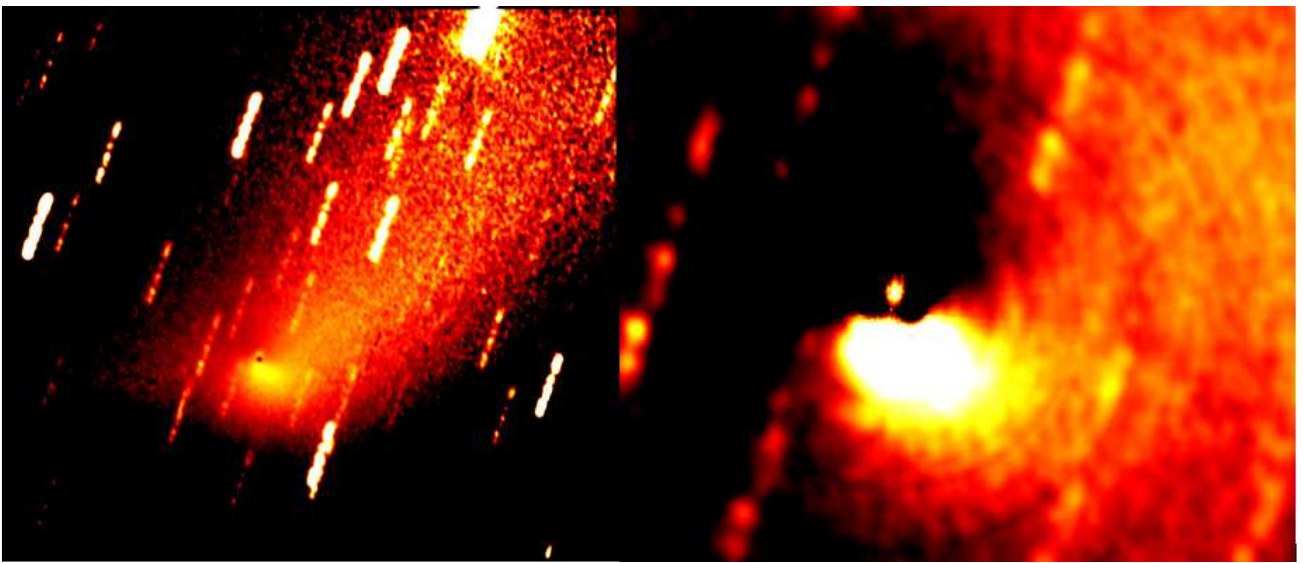

**Fig. 46 – 2013 May 11th, left RWM filter, right MCM filter.**

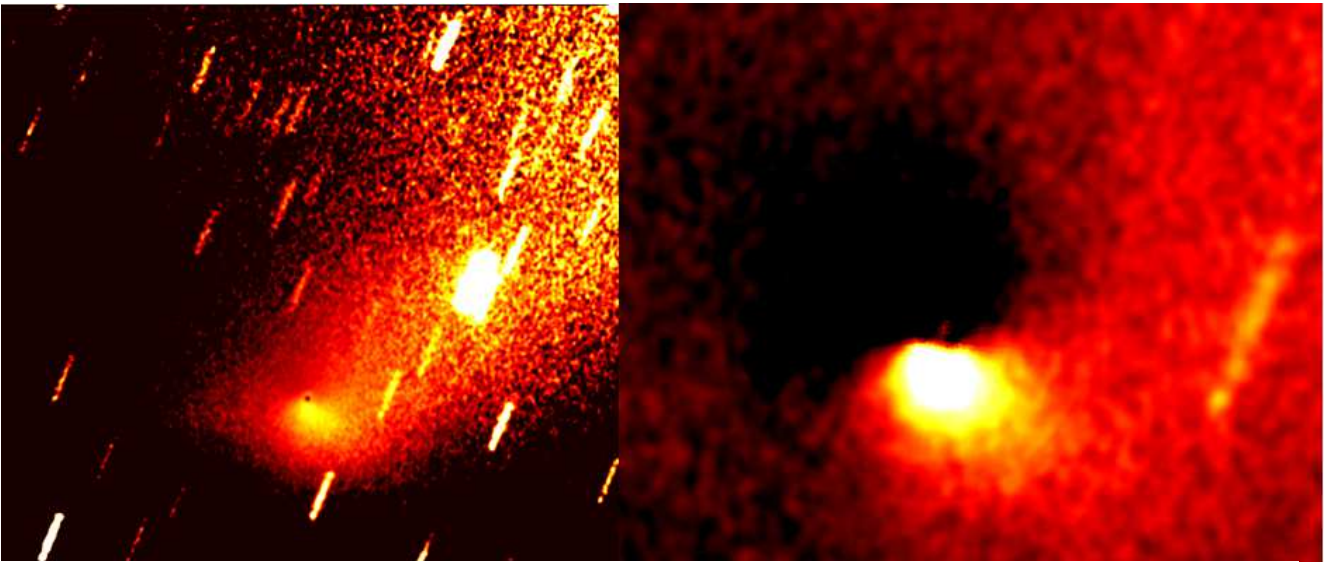
**Fig. 47 – 2013 May 14th, left RWM filter, right MCM filter.**

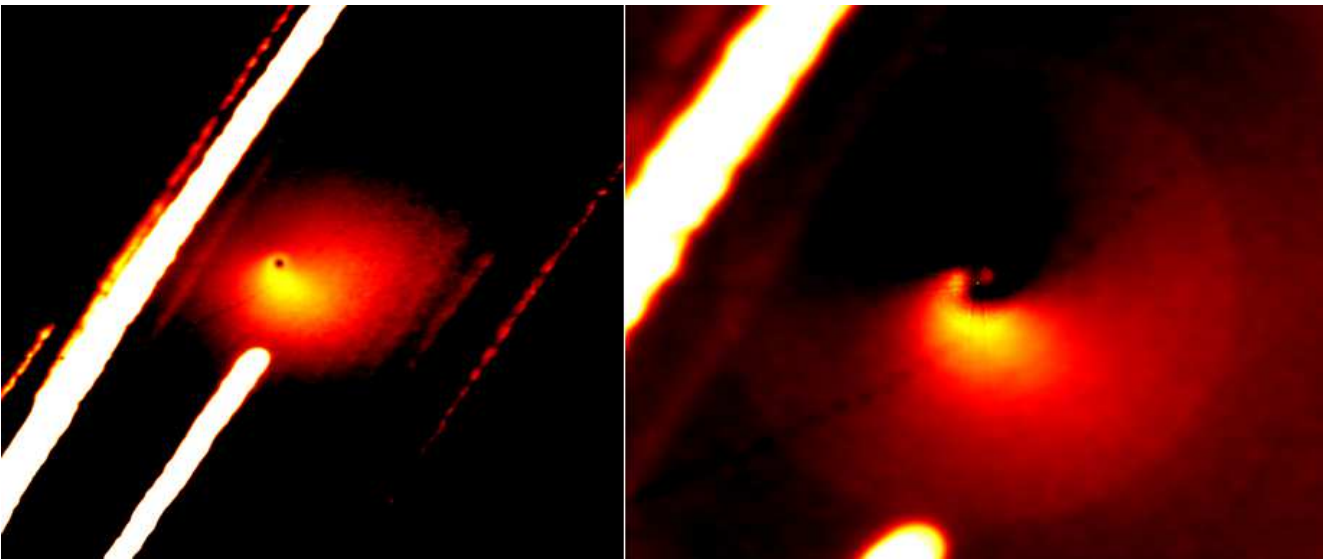
**Fig. 48 – 2013 May 17th, left RWM filter, right MCM filter.**

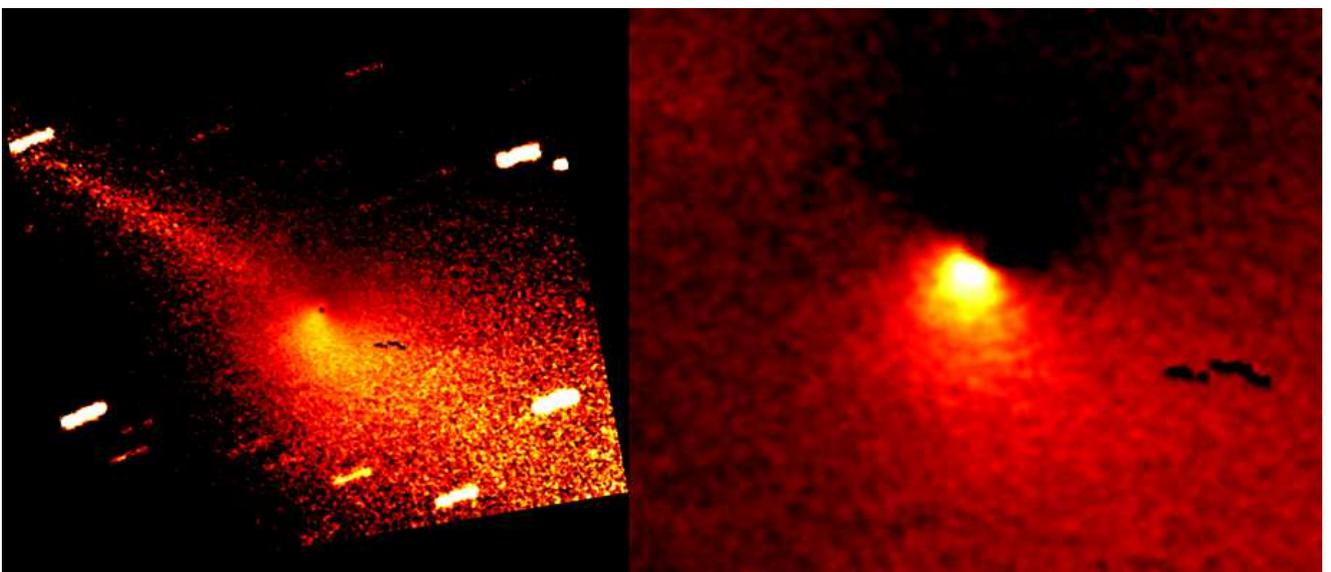
**Fig. 49 – 2013 May 25th, left RWM filter, right MCM filter. Note the anti-tail.**

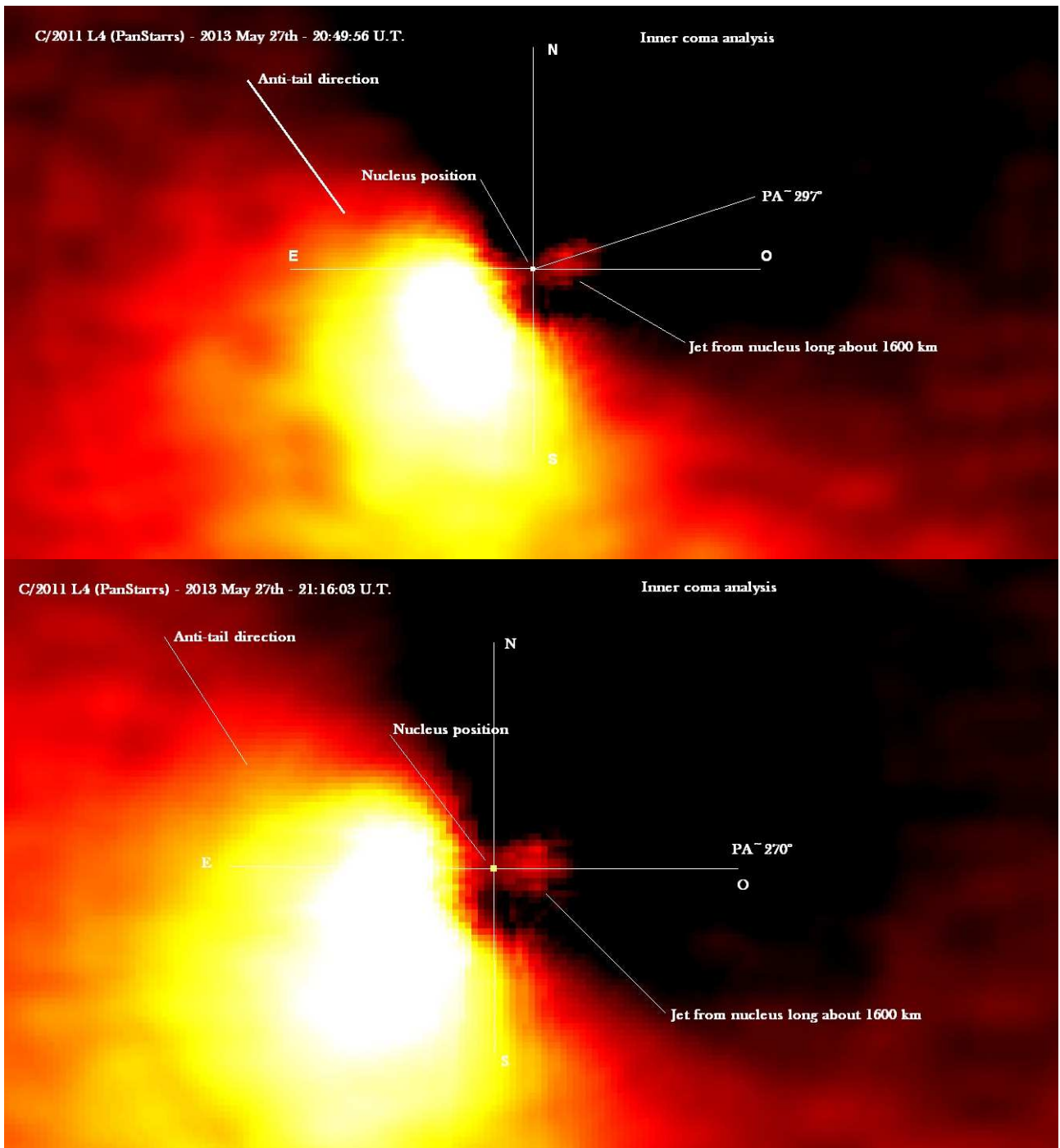

**Fig. 50 – 2013 May 27th, top 20:49:56 U.T. bottom 21:16:03 U.T., inner coma details extract with MCM model (Median Coma Model), show the "jet like" structure outgoing from the nucleus changing position with the time. The image was enhanced with the procedure described in the section 4.4 and 5. 2. Telescope 25 cm Newton f/4.8 and CCD Atik 16IC with Rc filter.**

The **"jet like"** structure outgoing from the nucleus, long about **1600 km**, clearly change position and in **26,11 min** rotated by an angle of **27°+/-3°.** Then, to rotate of 360° need **347,6+/-34,8 min**, equal to **5,79+/-0,58 hours**. This results is in agreement with the period found with the photometric analysis in the 4.3 section.

### 6) CONCLUSIONS

Here, were presented the results of the observations of the comet C/2011 L4 (PanStarrs) after perihelion. Comet PanStarrs, passed to the perihelion on 2013 March 10$^{th}$ at 0.3 A.U. showed a very strong dust production rate in the days following the perihelion passage. My measurement confirms that the comet is dust rich, have a small nucleus and rotate fast in about 5 hours. Much probably, the close passage to the Sun have produced a total sublimation of the ice on the surface and delivered a great quantity of dust. The fan shaped tail formed in the days immediately after the perihelion passage and the very long anti-tail observed in May month, confirms that a great quantity of dust was released from the nucleus. Af(rho) value has reached an impressive value of about 3000 m on 2013 April 6$^{th}$ . This value is not much precise because measured on images taken with CMOS Canon 10 D digital reflex, but it is surely a good indicator about the strong activity of comet PanStarrs. More precise measures have been obtained with CCD images filtered with narrow-band Rc photometric filter. Form 2013 April 16$^{th}$ to 2013 August 3th the comet moving away from the Sun, he has maintained a good rate of dust production. On 2013 August 3 to 2,8 A.U. form the Sun, Af(rho) value was about 40 m that, considering a linear relation between Af(rho) and Qdust rate, is equal to about 3,46x10^8 kg/day of dust lost. In my observation there are no signs of gas tail, so we can assume that the quantity of gas produced it was very low. Also the size of the comet is small, of radius about 500 m in average. Considering that come C/2011 L4 is a non-periodic comet from the Oort cloud, with a period of about 105000 years, computed after the perihelion, Epoch 2050, look like strange the fact that has resisted at the perihelion passage. Also if the perihelion distance is not too close, the rate of dust production after the perihelion is so great that we can assume that a good part of the surface has been eroded. Also, the period of rotation of about 5 hours has probably helped to expose, probably, the entire surface of the comet to solar heating for many times in a short period. We can therefore assume that the strong thermal and dynamic stress has produced the activation of many areas of the comet's surface. Surely, the detailed analysis of the inner coma show two "jet like" structures outgoing from the nucleus. In section 5 I measured the rotation of the less intense "jet" rotating in about 5 hours in agreement with result obtained by the photometric analysis on three days on July-August. Then, a small nucleus and fast rotator with an high rate of dust production is in perfect agreement with the big show that the non-periodic Oort cloud comet C/2011 L4 (PanStarrs) offered to Northern observers after perihelion passage as expected.

## Acknowledgements.
To me.